\documentclass[universe,review,accept,moreauthors,pdftex]{Definitions/mdpi} 

\usepackage{hyperref}
\usepackage{url}
\usepackage{verbatim}
\usepackage{natbib}

\firstpage{1} 
\pubvolume{1}
\issuenum{1}
\articlenumber{0}
\pubyear{2021}
\copyrightyear{2021}
\externaleditor{Academic Editor: {Tina Kahniashvili}} 
\datereceived{6 April 2021} 
\dateaccepted{3 June 2021} 
\datepublished{} 
\hreflink{https://doi.org/} 

\Title{Axion-Like Particle Searches with IACTs}
\TitleCitation{Axion-Like Particle Searches with IACTs}

\Author{Ivana 
 Batković $^{1,2,*}$ 
\orcidA{}, Alessandro De Angelis $^{1,2,3}$\orcidB{}, Michele Doro $^{1,2}$\orcidC{} and Marina Manganaro $^{4}$\orcidD{}

\AuthorNames{Ivana Batković, Alessandro De Angelis, Michele Doro and Marina Manganaro}
\AuthorCitation{Batković, I.; De Angelis, A.; Doro, M.; Manganaro, M.}
\address{%
$^{1}$ \quad  Università di Padova (UniPD), Dipartimento di Fisica e Astronomia (DFA) G. Galilei, I-35131 Padova, Italy; alessandro.deangelis@unipd.it (A.D.A.); michele.doro@unipd.it (M.D.) 
 \\
 $^{2}$ \quad Istituto di Fisica Nucleare (INFN), sez. Padova, I-35131 Padova, Italy; \\
 $^{3}$ \quad Instituto Superior Técnico, Universidade de Lisboa and LIP, Av. Rovisco Pais, 1, 1049-001 Lisboa\\
$^{4}$ \quad University of Rijeka, Department of Physics, 51000 Rijeka, Croatia; marina.manganaro@uniri.hr (M.M)}}

\corres{Correspondence: {ivana.batkovic@phd.unipd.it}}

\abstract{
The growing interest in axion-like particles (ALPs) stems from the fact that they provide successful theoretical explanations of physics phenomena, from the anomaly of the $CP$-symmetry conservation in strong interactions to the observation of an unexpectedly large TeV photon flux from astrophysical sources, at distances where the strong absorption by the intergalactic medium should make the signal very dim. In this latter condition, which is the focus of this review, a possible explanation is that TeV photons convert to ALPs in the presence of strong and/or extended magnetic fields, such as those in the core of galaxy clusters or around compact objects, or even those in the intergalactic space. This mixing affects the observed $\gamma$-ray spectrum of distant sources, either by signal recovery or the production of irregularities in the spectrum, called 'wiggles', according to the specific microscopic realization of the ALP and the ambient magnetic field at the source, and in the Milky Way, where ALPs may be converted back to $\gamma$ rays. ALPs are also proposed as candidate particles for the Dark Matter.
Imaging Atmospheric Cherenkov telescopes (IACTs) have the potential to detect the imprint of ALPs in the TeV spectrum from several classes of sources. In this contribution, we present the ALP case and review the past decade of searches for ALPs with this class of instruments.}


\begin{document}


\section{Axion and~Axion-Like-Particles}
\label{theory}
The presentation of a new fundamental particle called `axion' traces back to the late 1970s, when \citet{peccei:1977} introduced it as a possible solution to the otherwise-unexplained missing $CP$-simmetry violation in strong interactions. The~term `axion’ was first used by \citet{PhysRevLett.40.223}, who classified it as a “light, long-lived, pseudoscalar boson'' together with \citet{PhysRevLett.40.279}. Since then, axions were subject of strong scrutiny, from both theory and observation; however,~half a century later, they  remain one of the most compelling solutions to this so-called strong $CP$ problem.

Although there is nothing in the theory forbidding it, and,~therefore, it is expected, a~violation of the Charge $\times$ Parity $(CP)$ symmetry in Quantum Chromo-Dynamics $(QCD)$ was never experimentally observed. The~term of the  Lagrangian corresponding to $CP$ violation can be written as
\begin{equation}
     {\mathcal{L}}_{\theta_{QCD}}={{\theta}_{QCD}}\frac{g^{2}}{32 {\pi}^{2}}{G}^{a}_{\mu\nu} {\tilde{G}^{\mu\nu}_{a}},
     \label{eq:lagrangian_theta}
\end{equation}
where $\theta_{QCD}$ is a phase parameter of $QCD$, $G$ is the gluon field strength tensor, $a$ indicates trace summation over the $SU(3)$ colors and ${g}^{2}$ is the $QCD$ coupling constant. ${\theta_{QCD}}=0$  in case of no $CP$ violation.

For example, the~electrical dipole moment of the neutron, ${d}_{n}$, which shows a dependence on the $\theta_{QCD}$ angle, and~is, therefore, sensitive to the $CP$ violation term,  is experimentally bound  \citep{2020PhRvL.124h1803A}  to be
$        |{d}_{n}|\leq 1 \times {10}^{-26}  \:e\, \textup{cm}$,
which translates into $\theta_{QCD}<10^{-10}$, revealing a {\em fine-tuning} problem. 

The Peccei--Quinn (PQ) mechanism solves the Strong $CP$ Problem by introducing a new global symmetry, known as ${U(1)}_{PQ}$ symmetry, which makes the $CP$-violating term (Equation~(\ref{eq:lagrangian_theta})) in the $QCD$ Lagrangian negligible. Axions are, therefore, pseudo-Nambu--Goldstone bosons associated with the breaking of the ${U(1)}_{PQ}$ symmetry~\citep{PhysRevLett.40.223, PhysRevLett.40.279}.

In the PQ formalism, the~axion is a particle of mass ${m}_{a}$ and decay constant ${f}_{a}$, related to the decay amplitude, i.e.,~to the coupling.  In~the original model of the axion proposed by~\citet{peccei:1977, PhysRevLett.40.223, PhysRevLett.40.279}, the~axion decay constant ${f}_{a}$ is of the order of the electroweak scale ($\sim$246~GeV), and~the mass of the axion ${m}_{a}$ is inversely proportional to this. Its mass was, therefore, expected to be rather large, i.e.,~of the order of $100~$keV
\begin{equation}\label{eq:mass_sigma_a}
    {m}_{a}\simeq6\times{10}^{-6}~\textrm{eV}~\bigg(\frac{{10}^{12}~\textrm{GeV}}{{f}_{a}}\bigg) \, .
\end{equation}

Using experimental limits based on the stellar evolution and rare particle decays, this first model was ruled out. Soon after, two new models, abbreviated as $KSVZ$~\citep{PhysRevLett.43.103, 1980NuPhB.166..493S} and $DFSZ$~\citep{1981PhLB..104..199D, Zhitnitsky:1980tq}, emerged. They had in common the fact that the energy scale of the symmetry breaking was instead proposed to be large, i.e.,~close to the ``Grand Unification scale``, with the energy of ${10}^{15}$~GeV. This translated into a very light axion, with~mass ${m}_{a}\simeq{10}^{-9}~\textrm{eV}$. These axions would be very weakly coupled, hence the name currently used to dub them: ``invisible axions``. Taking into account their mass and coupling, these axions have eluded several experiments to date. 
{Furthermore, a similar but strictly massless pseudoscalar Goldstone particle was also considered, and named arion~\citep{ANSELM198239, 1988PhRvD..37.2001A}. }

\begin{figure}[H]
    \includegraphics[width=0.8\linewidth]{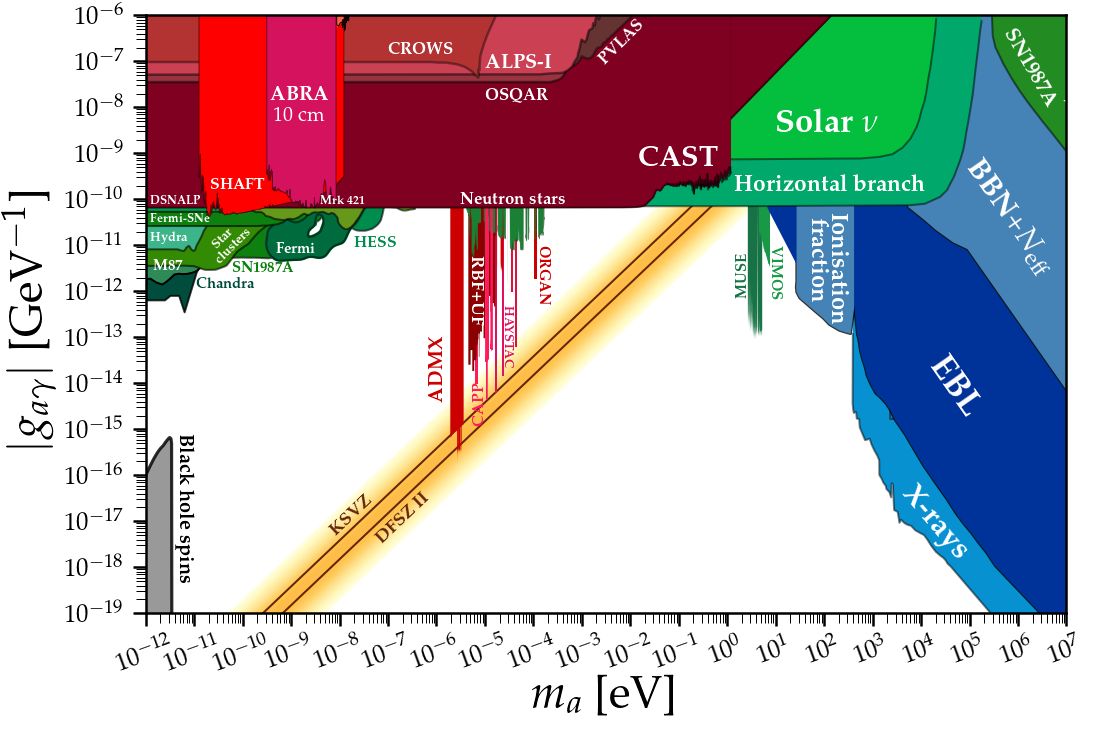}
    \caption{ALPs parameter space with current constraints (last update: July 2020). The collected limits, references and plots are available in the git-hub repository: \url{https://cajohare.github.io/AxionLimits/}  (accessed on June, 3rd 2020).} 
    \label{fig:ALPs_parameter_space}
\end{figure}

At present, after~many unsuccessful searches for axions (see Figure~\ref{fig:ALPs_parameter_space} for a collection of limits), the~axion model was extended to a wider group of particles, called Axion-Like Particles (ALPs), in~which the decay constant is no longer coupled with the axion mass, in~contrast with the original axion  (Equation~(\ref{eq:mass_sigma_a}))~\citep{2012JCAP...06..013A}. ALPs are also often found in SM extensions, motivated by string~theory. 

\begin{figure}[H]
    \includegraphics[scale=0.6]{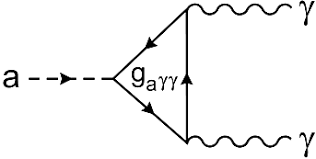}
    \caption{Feynman diagram of photon-axion coupling~vertex.}
    \label{fig:Feynman_diagram}
\end{figure}

A real ``treasure'' for the experimental detectability of ALPs is the term representing the axion coupling to photons through the two-photon vertex, shown in Figure~\ref{fig:Feynman_diagram}. The mentioned term is
\begin{equation}
         {\mathcal{L}_{{a}_{\gamma\gamma}}}=-\frac{{g}_{{a\gamma\gamma}}}{4}{F}_{\mu\nu}{\tilde{F}^{\mu\nu}}a={g}_{{a\gamma\gamma}}\vec{E}\cdot\vec{B}a,
         \label{eq:interaction_hunt_ALPs}
\end{equation}
where ${g}_{a\gamma\gamma}$ is the photon-ALP coupling, $F_{\mu\nu}$ the strength tensor of the electromagnetic field, ${\tilde{F}^{\mu\nu}}$ its dual, $a$ is the
axion field with mass ${m}_{a}$,~{$\vec{E}$ is the electric field of a beam photon, and~$\vec{B}$ is the external\footnote{By {external} magnetic field, we mean that the field is present outside the photon-ALP system itself and not generated during or by the interaction.}} magnetic field. This effect, explained as the photon-ALP conversion, occurs in  magnetic fields and is the basis of many experiments in the search for~ALPs. 

{More recently, axions were also proposed as viable Dark Matter (DM) particle} candidates. The~reason for this relies upon their small mass, combined with a possibly large decay constant ${f}_{a}\simeq{10}^{12}$~GeV. Since they are connected to spontaneous symmetry breaking, they could have been produced in the early Universe via ``misalignment” mechanisms. As~such, they could represent a substantial fraction of DM. \citet{2012JCAP...06..013A} report that, in order to explain the current amount of DM with ALPs, the~axion coupling, dependent on the mass of axions, has to be
\begin{equation}
    {g}_{a\gamma\gamma}<{10}^{-12}{\bigg[\frac{{m}_{a}}{1~\textrm{neV}}\bigg]}^{1/2}{\textrm{GeV}}^{-1}.
\end{equation} 
Together with hidden photons, axions pose as viable candidates for DM, and are named Very Weakly Interacting Slim Particles (WISPs).

\subsection*{Experimental Searches for ALPs}
\label{exp}
A wide class of axion searches are performed with special helioscopes, i.e.,~instruments pointing at the Sun, such as the well-known CERN Axion Solar Telescope (CAST)~\citep{2017NatPh..13..584A}. Axion helioscopes search for axions produced in the interior of the Sun by the conversion of plasma photons in the Coulomb field of charged particles, the~so-called Primakoff process. By~creating a strong magnetic field in the instrument and placing an X-ray detector at the far end, these detectors aim to reveal the reconversion of axions into X-ray photons~\citep{1983PhRvL..51.1415S}. CAST uses a dipole magnet with a strength of $\approx$9~T and length $L=9.26$~m. The~latest constraint on the coupling of photons to axions obtained with CAST~\citep{2017NatPh..13..584A} is 
$    {g}_{a\gamma\gamma}<6.6~\times~{10}^{-10}~{\textrm{GeV}}^{-1}$. 
Progress in this detection technique is expected from the new-generation axion helioscope International Axion Observatory (IAXO)~\citep{2015PhPro..61..193V}. 
Methods to constrain solar axions can be obtained using the M\``{o}ssbauer~\citep{2001PhRvD..64k5016K} and axioelectric effects~\citep{2017PhRvL.118z1301A}, among~others.

Alternative methods are pursued in so-called `light-shining-through-the-wall (LSW)' experiments, in~which photons from a strong laser beam are searched beyond a wall that can be crossed over by ALPs but not photons, asas ~wdone with The Optical Search for QED Vacuum Bifringence (OSQAR)~\citep{PhysRevD.92.092002} at CERN. There are also experiments based on the expected axion-induced birefringence of the vacuum, such as ALPS~\citep{2010PhLB..689..149E}. As~proposed by \citet{1983PhRvL..51.1415S}, another observable phenomenon could be the conversion of axions to photons in a resonant cavity. This study laid the theoretical ground for modern experiments such as Axion Dark Matter eXperiment (ADMX)~\citep{2018PhRvL.120o1301D}.

Astrophysics searches for axion and ALPs use cosmic magnetic fields and ample photon fluxes present in the cosmos. Clusters of galaxies, for~instance, have magnetic fields at their cores that are orders of magnitude larger than the average intracluster \mbox{medium~\citep{2008SSRv..134..311D,2003ApJ...590..225C,2006MNRAS.368.1500T,2012A&A...541A..99A,2017JCAP...12..036M}}. Magnetic fields in active galactic nuclei or pulsars could also be considered as a possible ``medium'' for the conversion of photons in axions or ALPs. In~the following section, we will focus on astrophysics experiments in the gamma-ray~range.

\section{Phenomenology of the Mixing between Gamma-Rays and ALP and Propagation in the Astrophysical~Environment}
\label{sec:pheno}

The existence of axions and ALPs can be probed by their imprints on the spectra of astrophysical sources. This is due to the fact that, in {the presence of} magnetic fields, ALPs couple with photons. Therefore, TeV gamma rays travelling over cosmological distances can oscillate to photons due to the interaction with magnetic fields, and/or convert to ALPs in strong magnetic fields and, as such, cross astrophysical distances until they possibly encounter another strong magnetic field, such as that of the Milky Way, in~which they can convert back into observable gamma rays. All these conversion/reconversion processes are governed by a probability term for the mixing $P_{\gamma\gamma}$, which depends on the actual ALP mass and coupling, as~well as the magnetic field~characteristics.

\subsection{ALP {Propagation}}In order to understand the phenomenon of conversion, it is necessary to compute the term $P_{\gamma\gamma}$. The~Lagrangian of the photon-ALP system can be written as
\end{paracol}
\nointerlineskip

\begin{equation}
\begin{split}
    \mathcal{L}=\frac{{g}_{{a}_{\gamma\gamma}}}{4}{F}_{\mu\nu}{\tilde{F}^{\mu\nu}}\:a -\frac{1}{4}{F}_{\mu\nu}{F}^{\mu\nu} + 
    \frac{{\alpha}^{2}}{90\:{m}_{e}^{4}}\bigg[({F}_{\mu\nu}{F}^{\mu\nu})^{2} + \frac{7}{4}{({F}_{\mu\nu}{\tilde{F}^{\mu\nu})}^{2}}\bigg]+ \frac{1}{2}(\partial_{\mu}a\:\partial^{\mu}a-{m}_{a}^{2}\:{a}^{2}),
    \end{split}
\end{equation}

\begin{paracol}{2}
\switchcolumn
\noindent where the first term relates to the photon--ALP coupling ${\mathcal{L}}_{a\gamma\gamma}$ term discussed in Equation~(\ref{eq:interaction_hunt_ALPs}), followed by terms relating to the effective Euler--Heisenberg Lagrangian ${\mathcal{L}}_{EH}$ for corrections of QED loops in photon propagators due to an external magnetic field~\citep{1988PhRvD..37.1237R}, where the last term ${\mathcal{L}}_{a}$ describes the kinetic and mass term of the axionic field. 
To model the propagation, we consider the motion of the ALP in the ${x}_{3}$ direction in a cold and ionized plasma. Generally, for~polarized photons and relativistic ALPs, the~equations of motion can be written as: 
\begin{gather}
    \bigg(i \frac{d}{d{x}_{3}} + E + \mathcal{M} \bigg) \begin{pmatrix}
    {A}_{1}({x}_{3})\\ 
    {A}_{2}({x}_{3})\\ 
    a ({x}_{3})
    \end{pmatrix}=0,
\end{gather}
where $\mathcal{M}$ is the photon-ALP mixing matrix. ${A}_{1}({x}_{3})$ and ${A}_{2}({x}_{3})$ represent the photon linear polarization amplitudes along the ${x}_{1}$ and ${x}_{3}$ axis, respectively, and~$a ({x}_{3})$ is the axion field strength \citep{2013PhRvD..87c5027M}. The~solution to this equation is the transfer function $\mathcal{T} ({x}_{3},0;E)$ using the condition $\mathcal{T}(0,0;E)=1$.

In case a homogeneous magnetic field transverse to the propagation direction (laying in ${x}_{2}$ direction) of the photon beam is assumed, then the photon--ALP mixing matrix $\mathcal{M}$ can be simplified into
\begin{gather}
    \mathcal{M}_{0}=
    \begin{pmatrix}
    {\Delta}_{\bot} & 0 & 0\\
    0 & {\Delta}_{\parallel} & {\Delta}_{a\gamma}\\
    0 & {\Delta}_{a\gamma} & {\Delta}_{a} 
    \end{pmatrix},
\end{gather}
where the elements in this matrix are written considering the plasma condition, the~QED vacuum birefringence effect, the~axion field, and~the photon--ALP mixing. They can be written as

\begin{gather*}
{\Delta}_{\bot}={\Delta}_{pl} + 2{\Delta}_{QED};\quad
\Delta_{\parallel}={\Delta}_{pl} + \frac{7}{2}{\Delta}_{QED};\quad 
    {\Delta}_{a\gamma} = \frac{1}{2} {g}_{a\gamma\gamma} {B}_{\bot}; \quad
    \Delta_{a}= - \frac{{m}_{a}^{2}}{2E}\\
\rm{with}\\
\Delta_{pl}= - \frac{{\omega}_{pl}^{2}}{2E} \quad \textrm{and} \quad {\Delta}_{QED}=\frac{\alpha E {B}_{\bot}^{2}}{45\pi {B}_{CR}^{2}}.
\end{gather*}

In the above equations, $\alpha$ is the fine structure constant and ${\omega}_{pl}$ is the plasma frequency, connected to the ambient thermal electron density, and~a critical magnetic field term ${B}_{CR}\sim4.4\times{10}^{13} \textrm{G}$ is defined. The~term ${\Delta}_{a\gamma}$ represents the photon-ALP mixing and depends on the strength of the interaction ${g}_{a\gamma\gamma}$ and the intensity of the transverse magnetic field ${B}_{\bot}$. Generally, the~magnetic field $B$ does not have to be in the ${x}_{2}$ direction, but~at an angle $\psi$ from it. In~this case, the~equations of motion are solved with a transfer function $     \mathcal{T}({x}_{3},0,E;\psi)=V(\psi)\:\mathcal{T}({x}_{3},0,E)\times {V}^{\dagger}(\psi)$ where $\mathcal{M}$ is changed in $\mathcal{M}=V(\psi)\:\mathcal{M}_{0}\:{V}^{\dagger}(\psi)$.

\subsection{Probability of ALP-Gamma Conversion}    \label{sec2.2}
With the transfer function, we can compute the probability of the conversion of a gamma ray to an ALP in an external magnetic field. The~ simplest description of the magnetic field is that of a single domain. In~this case, the~probability of the photon--ALP mixing can be written as \citep{1988PhRvD..37.1237R}
\begin{equation} 
\begin{aligned}
P_{\gamma \rightarrow a} &=\left(\Delta_{a\gamma}\: d\right)^{2} \frac{\sin ^{2}\left(\Delta_{\text {osc }} d / 2\right)}{\left(\Delta_{\text {osc }} \;d / 2\right)^{2}}
=\sin ^{2} (2 \theta) \sin ^{2}\left(\frac{\Delta_{\operatorname{osc}}d}{2}\right),
\end{aligned}
\label{eq:Pgammagamma}
\end{equation}
where $\theta$ is the rotation angle
$\theta=1/2\arcsin (2\Delta_{a\gamma} / {\Delta}_{osc})$, 
$d$ is the size of the domain and ${\Delta}_{osc}$ is the oscillation wave number,
${\Delta}_{osc}^{2}=[{({\Delta}_{a}-{\Delta}_{pl})}^{2}+4{\Delta}_{a\gamma}^{2}]$.
This term is often written in terms of a critical energy $E_{crit}$ defined as
\begin{equation}
{E}_{crit} \sim 2.5 \: 
\textrm{GeV}\:\frac{|{{m}_{a,neV}^{2}-{\omega}_{pl,neV}^{2}}|}{{g}_{11}{B}_{\mu G}},
\label{eq:E_crit}
\end{equation}
where ${\omega}_{pl,neV}$ is the plasma frequency in units of $\textrm{neV}$, $B_{\mu\textrm{G}}$ is magnetic field in microgauss and ${g}_{11}={g}_{a\gamma\gamma}/{10}^{-11}$~GeV$^{-1}$. The~critical energy is computed such that, around and above this value, the~probability of conversion $P_{\gamma \rightarrow a}$ in Equation~(\ref{eq:Pgammagamma}) becomes sizable. With~$E_{crit}$, the~term $\Delta_{osc}$ can be written as \citep{2012PhRvD..86g5024H} 
$
\Delta_{osc}=2 \Delta_{a \gamma} \sqrt{1+\left(E_{c}/E\right)^{2}}
$.

\subsection{Gamma-Ray Survival Probability}
We are now in the position to compute the gamma-ray survival probability, that is, the~fraction of photon that did \textit{not} convert to~ALP. 


To compute it, the~exact morphology of the magnetic field should be considered and the hypothesis of having just one single magnetic field domain with a fixed orientation is not plausible. A~common approach is to divide it into $N$ different domains. By~doing this, the~transfer matrix can be reformulated see \citep{2009JCAP...12..004M} properly, thus providing the total photon survival probability $P_{\gamma\gamma}$
\begin{equation}
P_{\gamma\gamma}=\frac{1}{3}\left(1-\exp \left(-\frac{3}{2} N P_{\gamma \rightarrow a}\right)\right).
\label{eq:final_pgg}
\end{equation}
When we write Equation~(\ref{eq:Pgammagamma}) following the previously introduced substitutions, we can obtain
\begin{equation}
   P_{\gamma \rightarrow a}=\sin ^{2} (2 \theta) \sin ^{2}\Bigg[\frac{{g}_{a\gamma\gamma}Bd}{2}\sqrt{1+\bigg({\frac{{E}_{c}}{E}\bigg)}^{2}} \Bigg].
   \label{eq:photon_survival_probability}
\end{equation}

As one can see from Equation~(\ref{eq:photon_survival_probability}), ${P}_{\gamma \rightarrow a}$ is dependent on the product of domain length $d$ and magnetic field $B$. Because~of this, it is essential to have a well-defined magnetic field model to account for the oscillations in the spectra of astrophysical objects caused by the photon-ALP~mixing.

\subsection{Astrophysical Magnetic Field and Photon Survival}
 $P_{\gamma\gamma}$ depends on the strength of the axion coupling to the photon, the~intensity, and~the coherence scale of the magnetic field in the medium in which the photon/ALP beam is propagating. While the first term is governed by the microscopic nature of ALP, there are several magnetic field realizations in the universe. Therefore, one needs to consider all different magnetic field environments in the path, from the source to the detector: photon-ALP mixing can be assumed in the magnetic field at the source, in~the local environment around the source, in~the intergalactic magnetic field and, finally, in the galactic magnetic field~\citep{2021arXiv210110270L}. Depending on the observed source, different combinations of magnetic fields can be considered, and~as reported by~\citet{1983PhRvL..51.1415S}, there are clearly several ways this problem can be approached. For~example, one of the most studied cases is that of the Active Galactic Nuclei (AGN) located in the cores of galaxy clusters. Here, once generated, gamma rays from the AGN would encounter the strong magnetic fields of the cluster core and have a sizeable chance of being converted to ALPs. Such an ALP could travel unimpeded along the intergalactic distances, whose magnetic field is extremely low,
 thus allowing only a moderate photon reconversion. Finally, the~ALP, when entering the Milky Way (MW) magnetic field, could (or could not) be reconverted back to gamma~rays.  

These are, therefore, several kinds of imprint in the original gamma-ray spectrum. In the first case, if~an ample fraction of photons is converted at the source into ALP that do not later convert back in the MW, a~signal depletion would be observed. In the second case: if an ample conversion happens in the source but then a back conversion happens in the MW, then one could also observe an ampler signal than expected; for~example, if~the ALP travelled regions of space that are opaque to gamma-rays (for example, regions with strong particles or radiation fields). One should mention that the above signatures would be observed on top of the well-known gamma-ray extinction due to the interaction with the Extragalactic Background Light (EBL)~\citep{2007PhRvD..76l1301D, 2008PhRvD..77f3001S, 2009PhRvD..79l3511S, 2011PhRvD..84j5030D,  2012PhRvD..86g5024H} which strongly limits the observation of TeV emission above redshift $z\sim 1$. The~propagation of VHE photons is affected by pair production processes with the EBL. Depending on the photon energy, they interact with the extragalactic background photons (EBL) or the cosmic microwave background (CMB), producing an electron--positron pair ($\gamma \to {e}^{+}+ {e}^{-}$). The~flux attenuation caused by these processes is  dominant for photon energies around $E_{\gamma}\approx~500$~GeV  and $E_{\gamma}\approx~{10}^{6}$~GeV, respectively \citep{2007PhRvD..76l1301D}. In~that way, the~greater part of photons is absorbed and evades detection: the universe becomes opaque to VHE gamma rays.
The above-mentioned cases of ALP signatures are possible in a regime above the critical energy $E_{crit}$ of Equation~(\ref{eq:E_crit}), where the photon--ALP mixing is maximum. A~third case is possible at around $E_{crit}$. In~this regime, the~oscillatory behaviour in Equation~(\ref{eq:photon_survival_probability}) would create 'wiggles' in the spectrum, in~correspondence with the probability term. These wiggles would be hardly misinterpreted as being of astrophysical origin and would, therefore, constitute a clear detection. Such a case is extensively discussed by, e.g.,~\citet{2007PhRvL..99w1102H, 2009MNRAS.394L..21D, 2009PhRvD..79l3511S, 2011PhRvD..84j5030D}.  
{\subsection{A Concrete Example of the Photon Survival Probability}}

As a showcase, in~Figure~\ref{fig:photon_survival_probability_gammaALPs}, we report the ${P}_{\gamma\gamma}$ calculated for ${m}_{a}=100~\textrm{neV}$ and ${g}_{a\gamma\gamma}= 1\times{10}^{-11}$GeV$^{-1}$, assuming conversion in the Perseus galaxy cluster magnetic field and in the Galactic magnetic field. The~reason for neglect of the intergalactic magnetic field in this case is its strength being restricted to $\textrm{B} \sim (0.1 - 1)~\textrm{neV}$, and~is still not confirmed. In~order to probe the photon--ALP conversions in a magnetic field of this strength, one needs to access critical energy ${E}_{crit}$ $\simeq500$~TeV or probe significantly low ALPs masses ${m}_{a}<{10}^{-10}$eV~\citep{2012PhRvD..86g5024H}.
On the other hand, there are works considering the photon--ALP mixing only in the host galaxy cluster magnetic field and the intergalactic magnetic field~\citep{2013PhRvD..88j2003A}, while some include all three of the mentioned magnetic field environments~\citep{2013PhRvD..87c5027M}.    
For the magnetic field of galaxy clusters, there are usually not well-established values, with the~bounds being between $10^{-15}$ G and $10^{-9}$ G, so their strengths are modeled assuming turbulence and using the Kolmogorov power spectrum. Regarding the magnetic field of the Milky Way, a few models are used most often
~\citep{2012ApJ...757...14J, 2008A&A...477..573S,2011ApJ...738..192P}. Most of these models are based on the Galactic Synchrotron Emission maps and the extragalactic rotation measurements, modelling a disk field and an extended halo field. In~one recent work~\citep{2020arXiv201108123D}, it is shown that, in ALPs searches using observations of BL Lacertae (often shortened as BL~Lac, a~well known blazar), there is a sizable jet-mixing effect, meaning that the modelling of the BL Lac jet magnetic field is needed. It is shown that the changes in the parameters of the jet model can cause changes in the photon--ALP mixing in a way that it will enlarge the part of ALP parameter space available for study. It is also shown that, in case of the sources embedded in strong cluster magnetic fields of dense environments, this effect is not relevant, so the constrains set by~\citet{2016PhRvL.116p1101A, 2013PhRvD..88j2003A} are still valid. In~the future, photon--ALP mixing in the blazar jet might become relevant and,~with the new generation of Cherenkov telescopes (\citep{Abdalla_2021}, e.g., the~Cherenkov Telescope Array (CTA)) the detection of more blazars at a higher redshift is expected. In~conclusion, very detailed magnetic field models are needed to address the photon-ALP mixing in a more accurate~way.
\begin{figure}[H]
    \includegraphics[scale=0.7]{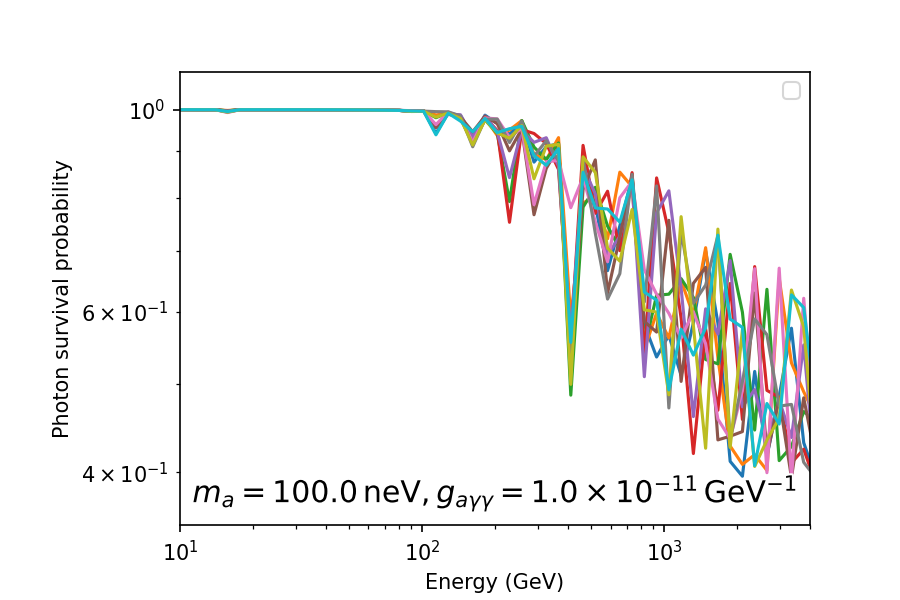}
    \caption{Photon survival probability for ${m}_{a}=100~\textrm{neV}$ and ${g}_{a\gamma\gamma}= 1\times{10}^{-11}$GeV$^{-1}$. Obtained using the GAMMAALPs code:~\url{https://github.com/me-manu/gammaALPs} (accessed on June, 3rd 2020).}
    \label{fig:photon_survival_probability_gammaALPs}
\end{figure}

\section{A Decade of Results with~IACTs}
\label{IACTs}
\unskip

\subsection{{VHE $\gamma$-ray Detection and Analysis Techniques}}
While there have been several early attempts to detect gamma rays at the ground starting, from the 1950s~\citep{HILLAS201319}, ground-based gamma-ray astronomy officially started with the detection, in 1989, of the Crab Nebula by the Whipple telescope, which has been operating since $1986$~\citep{1989ApJ...342..379W}. 
Increasingly new TeV emitters populated the gamma-ray sky, considering one of the last unexplored windows in the electromagnetic radiation from the cosmos. Whipple belongs to the Imaging atmospheric Cherenkov telescopes (IACTs) class. IACTs are suitable for the detection of VHE $\gamma$ rays, highly energetic photons which can be produced in the environments of astrophysical objects such as Active Galactic Nuclei (AGNs), supernovae, binary stars, pulsars, etc., as the result of highly accelerated (TeV-PeV) cosmic rays such as electrons and protons. The~sensitivity of IACTs is in the range $\sim$50 $\textrm{GeV}$--$50$ $\textrm{TeV}$. At present, Whipple is decommissioned, and~there are currently three major operating IACT arrays: the High Energy Stereoscopic System (H.E.S.S.)~\citep{2004NewAR..48..331H}, the~Major Atmospheric Gamma-ray Imaging Cherenkov Telescopes (MAGIC)~\citep{2001ICRC....7.2789L} and the Very Energetic Radiation Imaging Telescope Array System (VERITAS) \citep{2002APh....17..221W}.
{IACTs measure the energy and direction of $\gamma$-rays indirectly: when the $\gamma$-ray penetrates the atmosphere, it interacts with the present nuclei and produces a shower of particles. Charged particles belonging to the shower travel faster than the speed of light in the medium atmosphere, consequently producing Cherenkov light. The~faint Cherenkov light is collected by the mirror dishes of the telescopes and reflected into a camera positioned in front of the mirror dish. The~energy threshold of IACTs is inversely proportional to the signal-to-noise ratio, so it is convenient to maximize the mirror area and throughput of the optical system to  minimize the threshold. The~shape of the image shower is described by the so-called Hillas parameters~\cite{1985ICRC....3..445H}. The~flux in $\gamma$-rays is calculated using MonteCarlo simulations trained with OFF data, taking the collection area of the telescopes and the effective time of the observations into account. The~IACT observations are usually performed in the so-called wobble-mode, to~allow for the subtraction of the background during the observations~\cite{1994APh.....2..137F}. The~analysis of the data for the existing IACTs differs at the high level of analysis, when different methods to correct (unfold) the energy spectrum are used in the respective collaborations. The~unfolding methods can be based on different algorithms, in~order to assign to the $\gamma$-rays a true energy, and~to calculate the intrinsic spectrum of a source. In~particular, each array of IACTs possess a different configuration and asset so the instrument response function, used to obtain the final spectra, is different. The~principles of detection for IACTs are explained in detail in {Section} 2.2  of~\cite{2008RPPh...71i6901A}. At present, the collaborations are converging towards common software analysis tools, such as ctools\footnote{\url{http://cta.irap.omp.eu/ctools/} (accessed on June, 3rd 2020).}} and gammapy\footnote{\url{https://gammapy.org/}  (accessed on June, 3rd 2020).}. Despite a build-up of successes from the early Crab detection, the~technique became really mature in the first decade of this century, when not only were an increasing number of targets acquired, but~the results also reached a level of precision and significance never achieved before. As~an example, in~Reference~\cite{2015JHEAp...5...30A} MAGIC reports the spectrum of the Crab Nebula over three orders of magnitude in energy and four orders of magnitude in intensity, able to detect the source in less than 1~min. Along with this ramp-up of performance, the~attention moved from purely astrophysical interests to more fundamental questions, such as the possibility of observing the signature of ALPs in gamma-ray spectra. The~first decade of the 21st century brought interest in the imprints and modifications that the conversion of photons to ALPs and vice~versa could leave on the spectra of astrophysical objects~\citep{2007PhRvD..76b3001M, 2007PhRvD..76l3011H, 2007PhRvL..99w1102H, 2009PhRvD..79l3511S, 2009JCAP...12..004M, 2011PhRvD..84j5030D}. 
\subsection{{Astrophysical Targets for ALPs Searches with IACTs}}
{In the attempt to maximize the ALP signatures, it is possible to select the best target of observation. These are astrophysical emitters, where both ample, high-energy gamma-ray photons fluxes are produced, and~where the gamma-ray radiation encounters extended regions with significantly intense magnetic fields, which extend over much larger distances than their coherence length~\citep{2013PhRvD..88j2003A}. These conditions guarantee that the probability of interaction is maximal (see Equation~\eqref{eq:Pgammagamma}). Recently, \citet{Abdalla_2021} quantified the importance of the intensity of the magnetic field and the source brightness, showing that, for example, a~factor of 2.5 more intense magnetic field could result in factor 10 stronger constraints on the ALP coupling  (\citep{Abdalla_2021} Figure 7). In~the gamma-ray TeV sky, sources often display a flaring state, as opposed to a baseline emission state. If~possible, flaring states are then preferred to search for ALP. The~best candidates for observation are, therefore, Active Galactic Nuclei (AGNs), where particle acceleration and subsequent gamma-ray emission are found in the region around the central supermassive black holes (SMBHs). AGNs are the largest population of TeV targets. An~optimal situation is the AGNs being located in the central core of galaxy clusters, especially in a cool core one, in~which extended and intense magnetic fields permeate the region around the central galaxies. In~this condition, the~magnetic field is not only more intense (tens of
$\upmu$G) with respect to that in the intergalactic space, but~also more easily experimentally quantifiable. One of the best examples of this is the AGN NGC~1275 at the center of the Perseus Galaxy Cluster, presented above. Another class of objects of interest for ALP searches is compact objects, namely, pulsars and neutron stars, which are also present in binary systems. Here, the magnetic field is more localized, but significantly more intense. We will come back to discussion of source-specific information later in the text.}
{iI~order to make a prediction of the ALP--photon interaction pattern, one has to define both the microscopic nature of the ALP (mass and cross-section) as well as the magnetic field. 
For the former, one has to build a model of the interaction, as~done, for~example, in~the aforementioned, open-source gammaALPs code, and~scan the available parameter space. This is, at present, mostly done with grid sampling. 
For the magnetic field, since the knowledge is not accurate, the~procedure used is normally the computation of several random realizations and attempt of a marginalization procedure of the likelihood over this nuisance parameter, as~we will show later.}

{Other targets have been explored for ALPs searches. In~case of a supernova explosion, ALPs would be emitted via the Primakoff process and could be observed with $\gamma$ rays after a possible re-conversion in the magnetic field of the Milky Way. Following the observation of the supernova SN1987A, constraints due to the non-observation of $\gamma$ rays, coincidental with the neutrino observations, were set~\citep{1996PhLB..383..439B, 1996PhRvL..77.2372G}, but~affected by the strong uncertainties. Due to this,~\citet{2015JCAP...02..006P} revisited these papers using a more detailed analysis. Additionally, neutron stars are another possible candidate for ALP searches. Considering the radiative decays of axions produced by nucleon--nucleon bremsstrahlung in neutron stars,~\citep{1996slfp.book.....R, 2008LNP...741...51R},~\citet{2016PhRvD..93d5019B} have set constraints on the axion mass $m_a$ using the \textit{Fermi}-LAT data of four neutron stars.  This phenomenon was investigated in previous works with X-ray~\citep{PhysRevD.34.843} and $\gamma$-ray data from a supernova~\citep{2011JCAP...01..015G}. }

\subsection{{Critical Energy and Parameter Space for $\gamma$-ray Studies}}

This interest in ALP searches in the $\gamma$-ray range was firstly encouraged by the unexplained observation of a change in light polarization in a vacuum filled with a magnetic field detected by the Polarization of the Vacuum with Laser (PVLAS) experiment, \mbox{\citet{2006PhRvL..96k0406Z}}, that offered an explanation based on the existence of a light axion. The~results of the PVLAS experiment were in tension with the  astrophysical limits. In~order to reconcile the signal obtained with PVLAS, authors theorized an ALP with mass $m_{a}=1.3$~meV and coupling $g_{a\gamma\gamma}=3\times{10}^{-6}$~Gev$^{-1}$. Following on this interpretation, \citet{2007PhRvD..76b3001M} included photon--ALP conversion in the magnetic field of our galaxy and, taking the mentioned parameters into account, present the possible distortions in the photon spectra above the energies $E_{\gamma}\geq~10$~TeV. A~few months later,~\citet{2008PhLB..659..847D} and \citet{2007PhRvL..99w1102H} extended this approach. Taking into account the possibility of the photon--ALP conversion in and around the gamma-ray source, as~strong astrophysical accelerators, they showed that the critical energy in Equation~(\ref{eq:E_crit}) falls directly in the gamma-ray range. The~photon--ALP conversion then depends on the condition
\begin{equation}
    g_{a\gamma\gamma}B\,s/2\geq1
    \label{eq:condition_Hillas}
\end{equation}
where $B$ is the magnetic field component aligned with the photon polarization vector and $s$ is the size of the magnetic field domain. 
If the photon--axion conversion happens at the source, the~product $B\,s$ in Equation~(\ref{eq:condition_Hillas}) is directly connected to the Hillas criterion~\citep{1984ARA&A..22..425H} for the maximum possible acceleration energy of cosmic rays, and~taking into account that cosmic rays with energies up to a few times $10^{20}~$eV have been observed, it follows that sources with $B_G\,s_{pc}\geq0.3$ should exist \citep{2007PhRvL..99w1102H}. \citet{2007PhRvL..99w1102H} showed that IACTs such as H.E.S.S., MAGIC and VERITAS could have probed the range of masses of $m_{a}=(10^{-9}-10^{-3})$~eV with sensitivities stronger than CAST, as~shown in Figure~\ref{fig:Gamma_ALPs_parameter_space}. The~best candidates for observation were identified with AGNs located in the cores of galaxy clusters. One can now compare Figure~\ref{fig:Gamma_ALPs_parameter_space} with Figure~\ref{fig:ALPs_parameter_space} to see how \citet{2007PhRvL..99w1102H} were right in their predictions.

\begin{figure}[H]
    \includegraphics[scale=0.40]{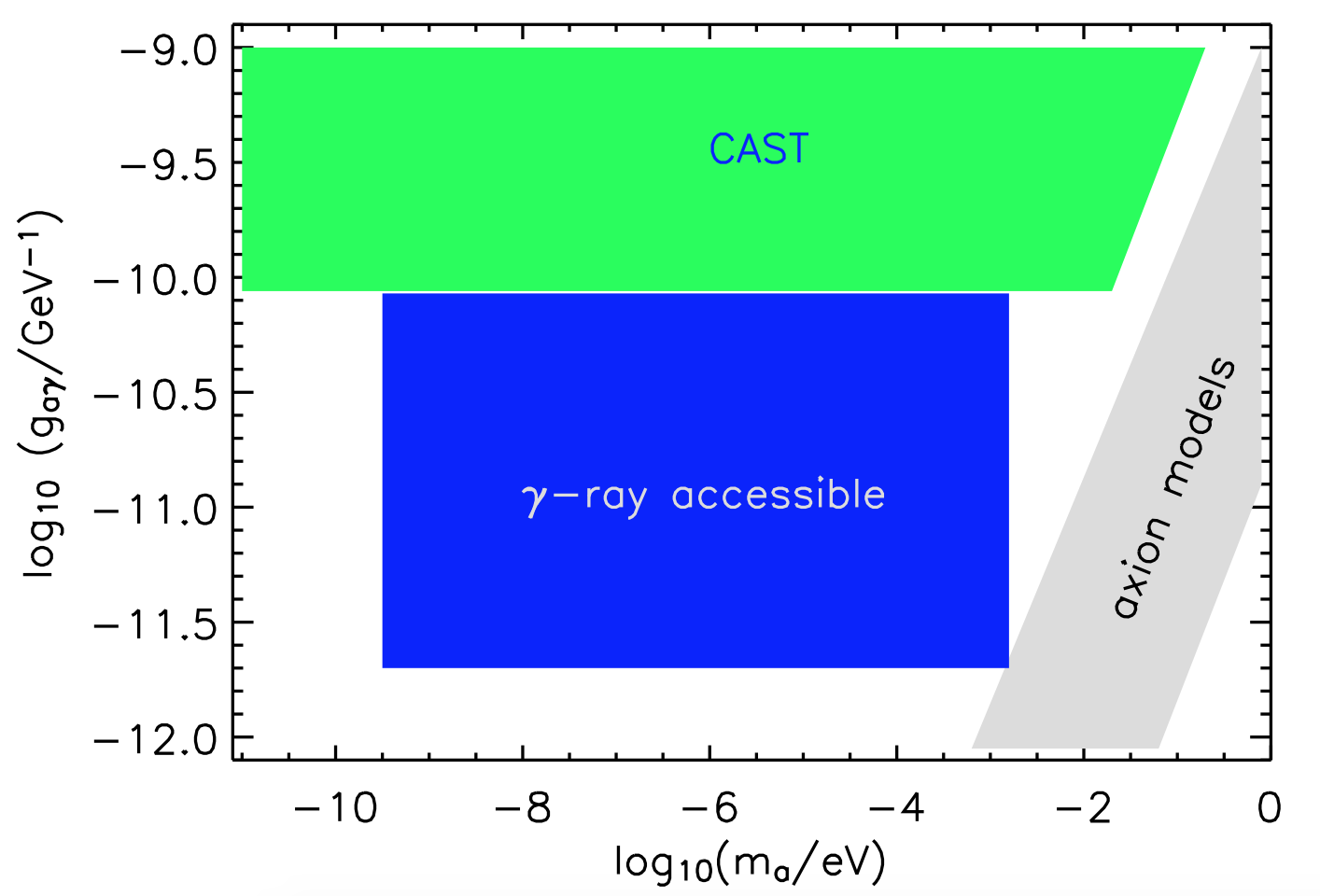}
    \caption{ALPs parameter space available for gamma-ray observations. Reprinted from~\citet{2007PhRvL..99w1102H}.}
    \label{fig:Gamma_ALPs_parameter_space}
\end{figure}
The first works by~\citet{2008PhLB..659..847D,2007PhRvD..76l1301D} are based on the unexpected  transparency of the universe: EBL observations at the time showed higher transparency at higher redshifts than expected \citep{2000PhLB..493....1P,2008Sci...320.1752M}. Following the idea that, if~converted to ALPs, photons could travel through the extragalactic space without interaction with the EBL or CMB photons, be converted back to photons in the Galactic magnetic field and be detected as such, the~photon--ALP conversion could reduce the opacity of the universe to VHE gamma rays, as~discussed above. In~order to explain the possible detection of TeV photons from a source located at $z=0.44$, which was not expected by conventional physics of photon propagation at the time, \citet{2009PhRvD..79l3511S} laid out a similar model. They built a model combining both the mixing near or in the source and mixing in the intergalactic space, stressing the importance of observations, both in the lower and highest energies in order to better constrain the intrinsic spectra of the sources, the~EBL attenuation and explore the morphology of the considered magnetic fields. The~photon flux attenuation was investigated by varying and combining the photon energy, magnetic field intensity, source redshift and ALPs parameters, showing that these effects could be observed in the spectra of AGNs at the higher energies, $E_{\gamma}\geq1$~TeV, especially if combined with the \textit{Fermi}-LAT energy regime \citep{2009PhRvD..79l3511S}.
After MAGIC detected the surprising rapidly varying emission from the flat spectrum radio quasar (FSRQ) PKS 1222+216~\citep{2011ApJ...730L...8A},~\mbox{\citet{2012PhRvD..86h5036T}} performed a combined ALPs study using the MAGIC and \textit{Fermi}-LAT data. The~aim of~\citep{2012PhRvD..86h5036T} was to present the emission model, including the photon-ALP oscillations mechanism, and explain the mentioned detection. The~results showed an agreement with the previously introduced De Angelis, Roncadelli and Mansutti (DARMA) scenario that includes photon--ALP oscillations triggered by large-scale magnetic fields to effectively reduce the EBL attenuation at the energies above $100$~GeV~\citep{2009MNRAS.394L..21D, 2011PhRvD..84j5030D}. These results showed the possibility of explaining such emissions with photon--ALPs oscillations by applying them to the other detected FSRQs.

{
The challenge related to the detection of spectral features induced by ALPs in the gamma-ray spectra is due to the the number of statistical and systematics fluctuations that shape the spectrum, even in the case of no ALP effect. First of all, the~intrinsic spectrum is shaped by the absorption by the EBL, as~discussed above. Such an effect is not-negligible for targets farther than $z\sim0.1$, but many models have been created based on EBL observations in the UV-infrared. Therefore, it is possible to correct the spectra for EBL absorption at different redshifts. The effects of LIV on the flux in photons could also compete with ALPs conversion, but~the power of a given source to constrain LIV increases with its distance, its variability in time and the hardness of its energy spectrum, so not all the considered targets are also good targets for studying LIV. The~energy reconstruction is generally performed  with IACTs at about 10--20\% precision, depending on the energy. Finally, the~data are affected by a variety of systematics due to the instrument itself (e.g., telescope mirror reflectivity) as well as external factors (atmospheric optical depth). While the former are estimated more accurately, less accurate results were obtained for the latter. 
Observing irregularities in the spectrum, such as~those caused by the ALPs---see Figure~\ref{fig:photon_survival_probability_gammaALPs}---is, therefore, challenging.}

\subsection{{H.E.S.S. Results with PKS~2155-304}}
After the first predictions of~\citet{2007PhRvL..99w1102H}, one of the first attempts to constraint ALP with gamma rays was made by H.E.S.S., using the data from the BL~Lac object PKS~2155-304 \citep{2013PhRvD..88j2003A}. In~this work, a~search for irregularities induced by the photon--ALP mixing in the spectrum was performed, and schematically shown in Figure~\ref{fig:spectral_irregularity_scheme}. 
{
The problem is in searching for ALP-induced spectral patterns on top of a spectrum generated by the main astrophysical processes at the source. Normally, these generate rather smooth and featureless spectra, such as power-laws, with~or without a cutoff or log-parabolic shape. \citet{2013PhRvD..88j2003A} assumed a power-law function as a local spectral model, justified by the processes explaining the acceleration and radiation in the extreme astrophysical sources, such as BL Lacs~\citep{2001A&A...367..809K}. For~an estimation of the irregularities,~\citet{2012PhRvD..86d3005W} proposed a reduced $\chi^2$ test with the null hypothesis build without the ALP ($\phi_{w/oALP}(\vec{\theta})$):}
\begin{equation}
   I=\frac{1}{d}\sum_{k}^{N}{\frac{(\phi_{w/oALP}(\vec{\theta})-\phi_{k})}{\sigma_{k}^{2}}}^{2}=\frac{{\chi}^{2}}{d},
   \label{eq:reduced_chi_square}
\end{equation}
{where $d$ is the number of degrees of freedom, $k$ runs over the $N$ bins, and~$\phi_{w/oALP}(\vec{\theta})$ is a global fit without ALPs with spectral parameters $\vec{\theta}$. This method relies on the accuracy of the assumed shape of the spectrum, and~is, therefore, subject to possible bias, but~can be used in the case when the global fit represents a good estimate on the spectrum~\citep{2020arXiv201012396C}. Expanding on this, \citet{2013PhRvD..88j2003A} searched for irregularities avoiding a global fit and using only a spectral shape over three adjacent points in the energy spectrum (a \mbox{triplet $i$}):}
\begin{equation}
    {\mathcal{I}}^{2}=\sum_{i}\frac{{\big(\tilde{{\phi}_{i}}-{\phi}_{i}\big)}^{2}}{{\vec{d}}_{i}^{T}C_{i}\vec{d}_{i}},
    \label{eq:estimator_hess}
\end{equation}
{where ${(\tilde{{\phi}_{i}}-{\phi}_{i}\big)}^{2}$ is the residual of the middle bin in the triplet, $\phi_i$ the measured flux, $\tilde{{\phi}_{i}}$ the flux in the median bin expected from the power-law fit to the side bins, $C_{i}$ covariance matrix for the triplet and ${\vec{d}}_{i}^{T}=\big(\frac{\partial\tilde{{\phi}_{i}}}{\partial{\phi}_{i-1}},-1,\frac{\partial\tilde{{\phi}_{i}}}{\partial{\phi}_{i+1}}\big)$.}
Although both methods showed consistent results, \citet{2013PhRvD..88j2003A} evaluated that, due to its independence of the global spectral model assumption, the sum of residuals over three adjacent spectral bins is preferred for this kind of analysis. This estimator is calculated for each set of ALPs parameters and 1000 spectra are simulated in order to take the randomness of both the intergalactic magnetic field and the galaxy cluster magnetic field into account. The distribution of values of the spectral irregularity estimator for both the observed spectrum and spectra folded with photon--ALP oscillations for different ALPs parameters are compared, and exclusions of the ALPs parameter space were obtained at 95~$\%$ confidence level. The~results (Figure~\ref{fig:spectral_irregularity_scheme}, right) yielded constraints on the photon--ALP coupling value ${g}_{a\gamma\gamma}<2.1\times{10}^{-11}~\textrm{GeV}^{-1}$ for masses of the ALPs ${m}_{a}$ in the range (15--60)~neV~\citep{2013PhRvD..88j2003A}. 
\end{paracol}
\clearpage
\nointerlineskip

\begin{figure}[H]
\widefigure
    \includegraphics[height=5.5cm]{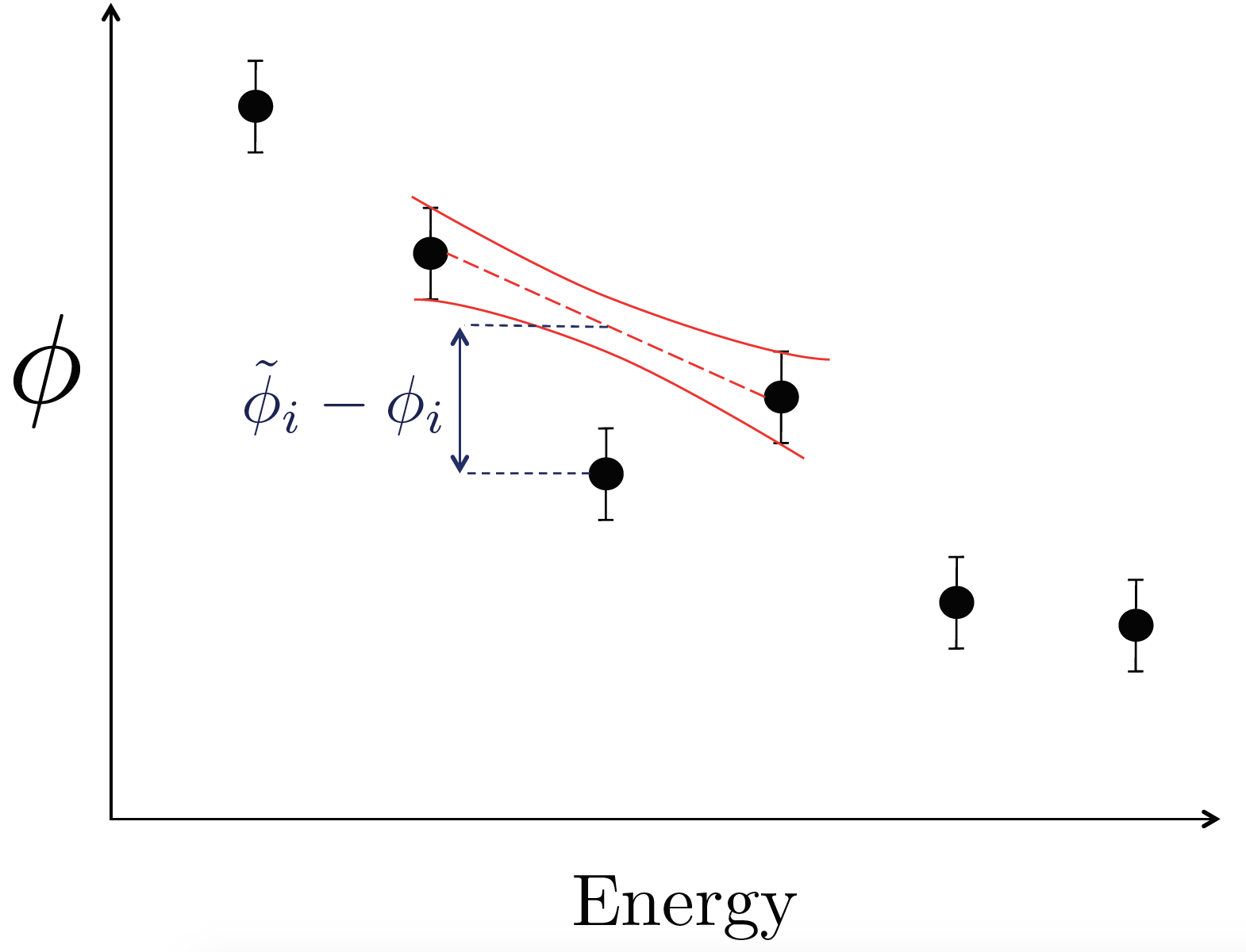}
    \includegraphics[height=5.5cm]{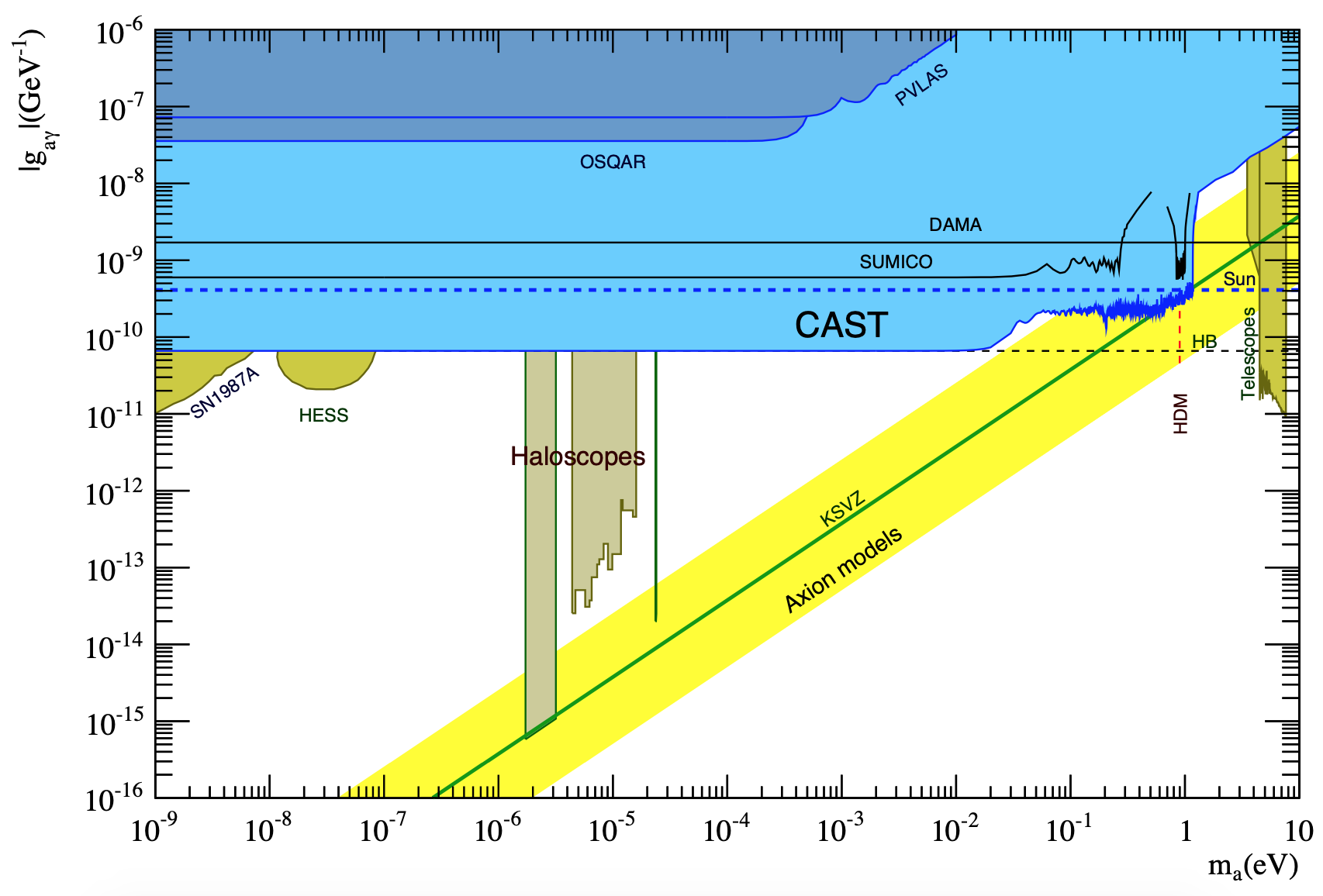}
    \caption{(\textbf{Left}) Schematic
 view of spectral irregularity quantification. Reprinted from~\citet{2013PhRvD..88j2003A}. (\textbf{Right}) Constraints on ALPs parameter space set by CAST, compared with results from the previous helioscope Sumico and DAMA experiment, as~well as with PVLAS~\citep{2016EPJC...76...24D} and OSQAR~\citep{PhysRevD.92.092002} experiments, constraints set by H.E.S.S collaboration, observations of SN1987A, Solar astrophysics and Dark Matter (DM) searches. Reprinted from \citet{2017NatPh..13..584A}.}
    \label{fig:spectral_irregularity_scheme}
\end{figure}

\begin{figure}[H]
\widefigure
    \includegraphics[height=5cm]{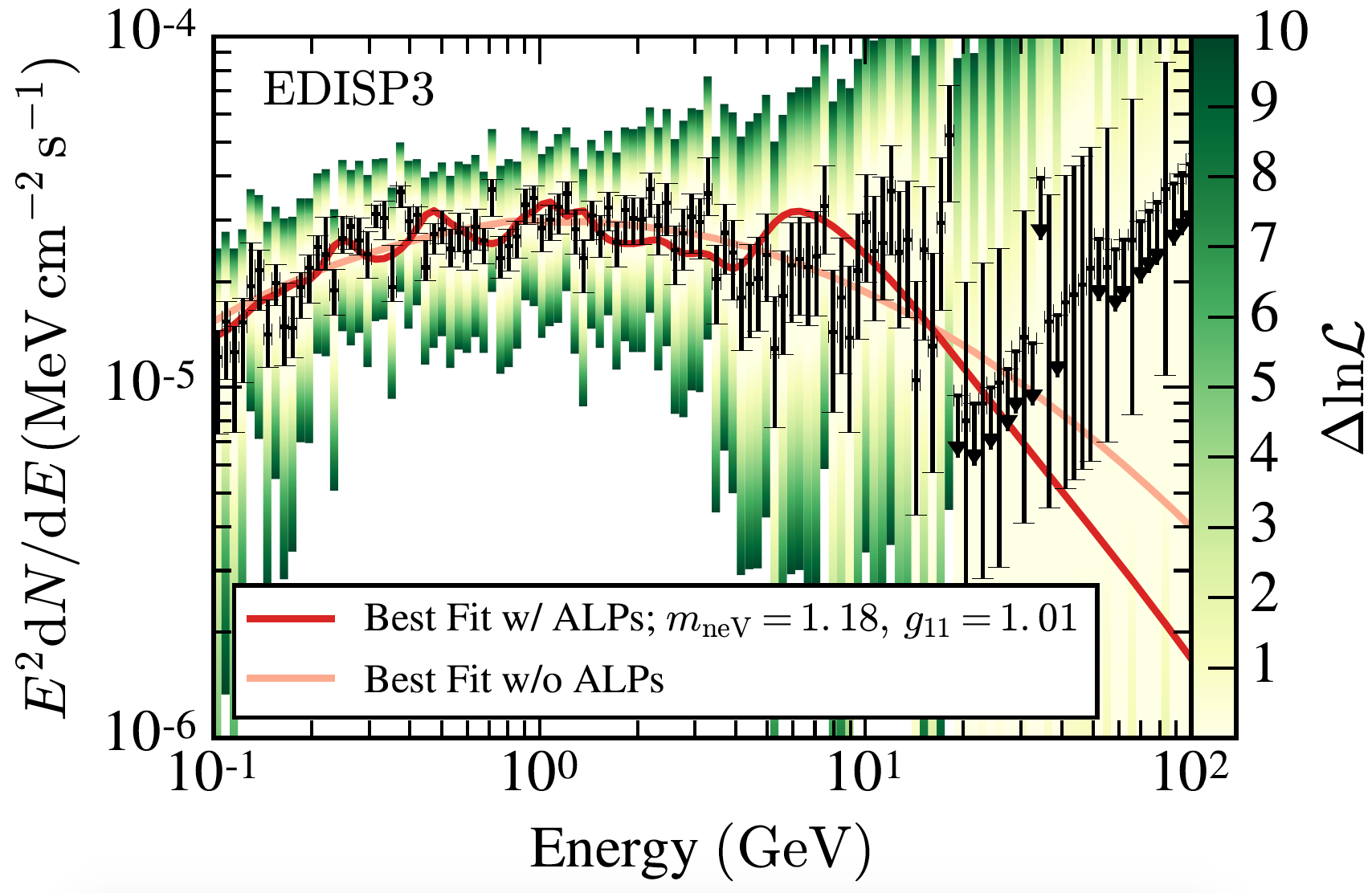}
   \includegraphics[height=5cm]{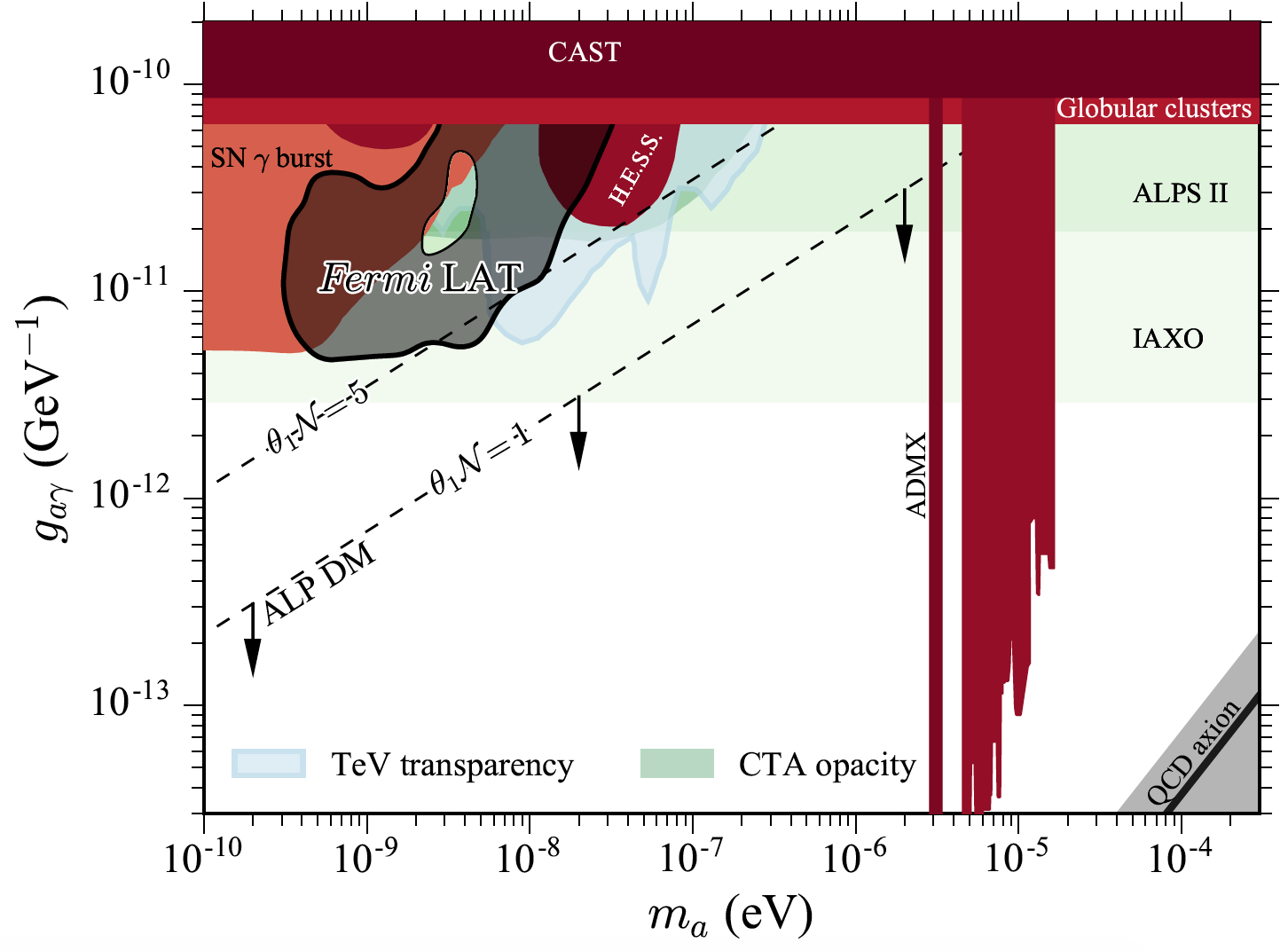}
    \caption{(\textbf{left}) Likelihood curves for one event type and best spectral fits with and without ALPs. Reprinted from~\citet{2016PhRvL.116p1101A}. (\textbf{right}) Projected limits on the ALPs parameter space obtained with the \textit{Fermi}-LAT study of the NGC~1275 data, compared with the results from other experiments at the time. Reprinted from~\citet{2016PhRvL.116p1101A}.}
    \label{fig:likelihood_curves_NGC1275}
\end{figure}

\begin{paracol}{2}
\switchcolumn

\vspace{-6pt}

\subsection{{Studies on Spectral Irregularities of NGC~1275}}
The IACT results were completed at lower energies, making use of the \textit{Fermi}-LAT instrument data. \citet{2016PhRvL.116p1101A} analyzed $6$ years of NGC~1275 data, collected with \textit{Fermi}-LAT, using the \texttt{Pass 8} event analysis, and produced ALP predictions by including the photon--ALP conversion in the intracluster magnetic field and in the galactic magnetic field of the Milky Way. A~fit of the time-averaged spectrum of NGC~1275 and ALPs models was made, and a likelihood analysis was performed. In~Figure~\ref{fig:likelihood_curves_NGC1275}, one can see the likelihood of one of the event types, together with the best spectral fit with and without ALPs. 
{
To evaluate the ALPs hypothesis, \citet{2016PhRvL.116p1101A} exploited a likelihood ratio test statistics ($TS$). In~the procedure, a~time-averaged spectrum is modelled by a smooth function, and likelihood is extracted for each reconstructed energy bin $k'$, $\mathcal{L}({\mu}_{k'},{\theta}|{D}_{k'})$, where ${\mu}_{k'}$ is the expected number of photons in the photon--ALP conversion scenario, ${\theta}$ are the nuisance parameters of the fit, and ${D}_{k'}$ is the observed photon count. For~each set of ALPs parameters and magnetic field, the~joint likelihood of all reconstructed energy bins $k'$ is maximized and the best-fit parameters are determined. Among~the different turbulent magnetic field realizations, simulated by accounting for its randomness, the~one corresponding to the 0.95 quantile of the likelihood distribution is chosen. The~likelihood ratio test is performed~as
\begin{equation}
    TS=-2~ln\Bigg(\frac{\mathcal{L}({\mu}_{0},\hat{\hat{\theta}}|{D})}{\mathcal{L}(\hat{\mu}_{95},\hat{\theta}|{D})}\Bigg),
\end{equation}
where the null hypothesis is the no-ALP scenario (including the EBL attenuation) with expected photon count ${\mu}_{0}$ and nuisance parameters $\hat{\hat{\theta}}$, and~the alternative hypothesis of ALP, shows an expected photon count ${\mu}_{95}$ and nuisance parameters $\hat{\theta}$~\citep{2016PhRvL.116p1101A}. Aside from the degeneracy of the photon--ALP conversion in coupling and magnetic fields, and~non-linearly scaled irregularities considering the ALPs parameters, in~comparison with the ALP hypothesis, the~null-hypothesis is independent of the realisations of the magnetic field. Considering this, the~null distribution needs to be derived from Monte Carlo  simulations~\citep{2016PhRvL.116p1101A}. The~exclusion threshold value, above~which the set of ALPs parameters can be excluded with the $95\%$ confidence level statistics, is also calculated from Monte Carlo simulations.}
The result of this research was the exclusion of the ALP coupling values in the range $0.5\times10^{-11}~\textrm{GeV}^{-1}\leq{g}_{a\gamma\gamma}\leq\times10^{-11}~\textrm{GeV}^{-1}$ for  ALPs masses $0.5~\textrm{neV}\leq{m}_{a}\leq5~\textrm{neV}$ and $g_{a\gamma\gamma}\geq~1\times10^{-11}~\textrm{GeV}^{-1}$ for $5~\textrm{neV}\leq{m}_{a}\leq10~\textrm{neV}$~\citep{2016PhRvL.116p1101A}, as~seen in Figure~\ref{fig:likelihood_curves_NGC1275}.  
\end{paracol}
\nointerlineskip

\begin{figure}[H]
\widefigure
    \includegraphics[height=5.5cm]{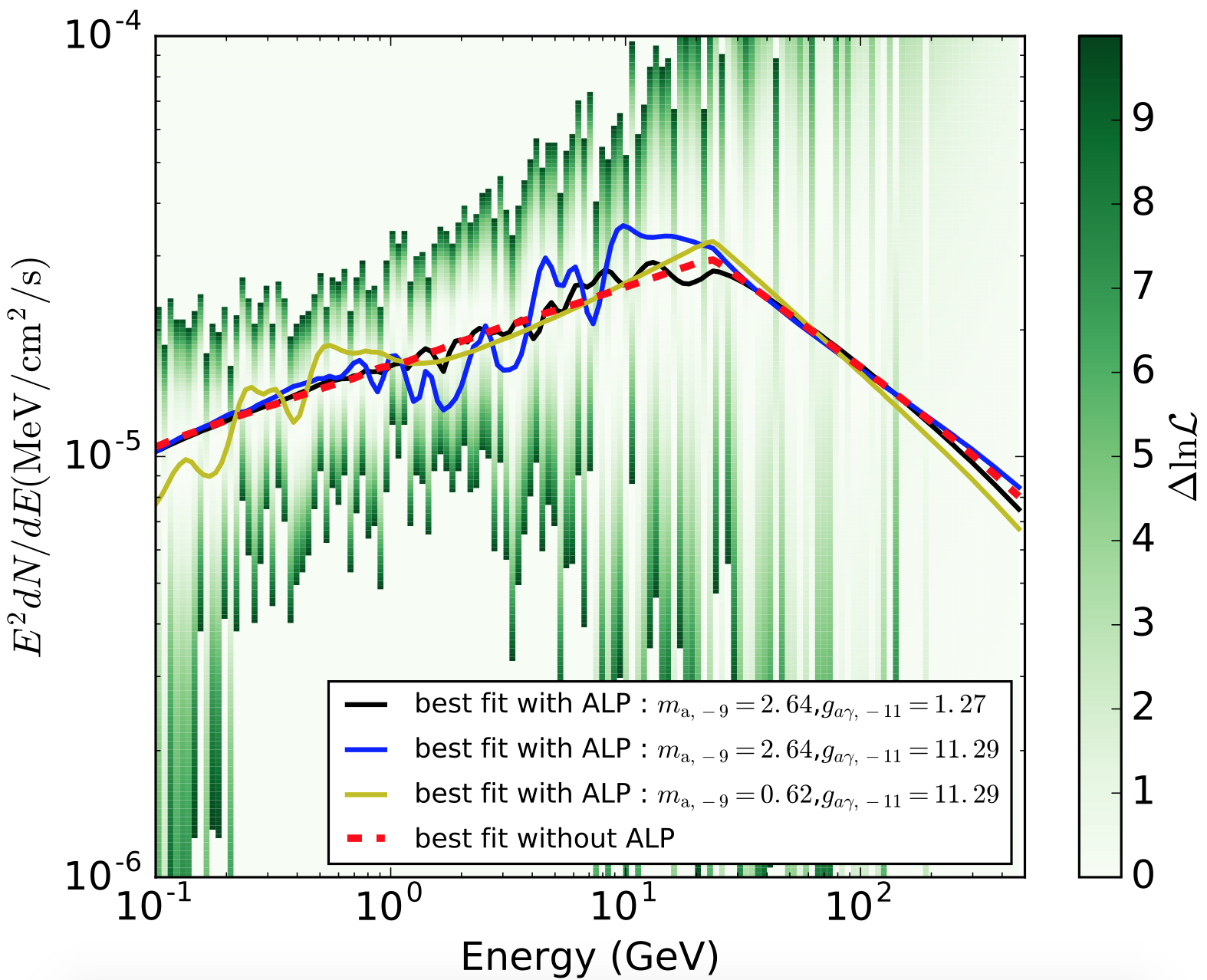}
    \includegraphics[height=5.5cm]{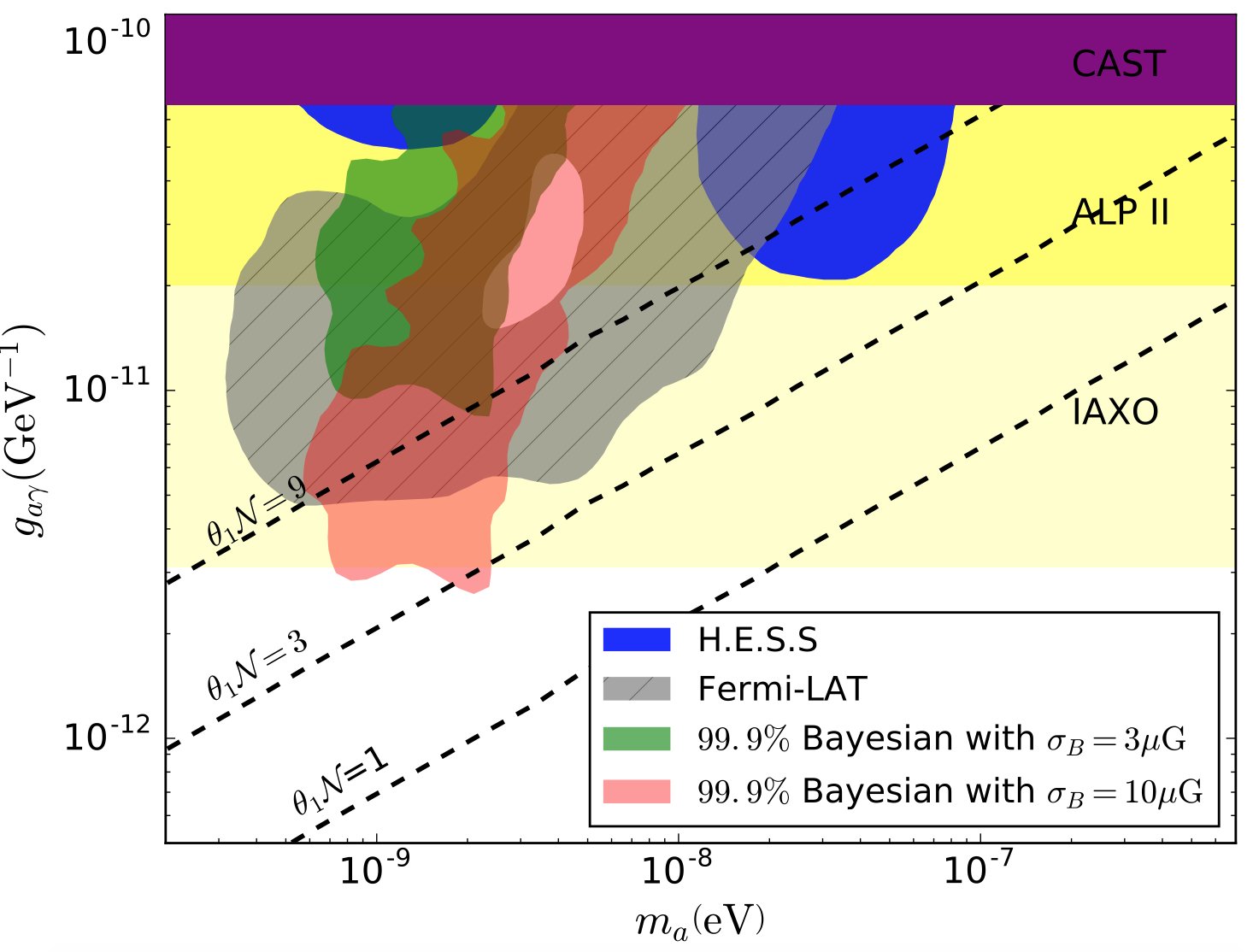}
    \caption{(\textbf{left}) Likelihood curves for the observed spectrum of PKS 2155-304. Solid lines represent, best fits including the photon--ALP oscillations and best spectral fit without oscillations included. Reprinted from~\citet{2018PhRvD..97f3009Z}. (\textbf{right}) Comparison of exclusion regions derived in \citep{2018PhRvD..97f3009Z}, compared with exclusion regions from H.E.S.S. observations of PKS-2155-304~\citep{2013PhRvD..88j2003A} and \textit{Fermi}-LAT observations of NGC~1275~\citep{2016PhRvL.116p1101A}. Reprinted from~\citet{2018PhRvD..97f3009Z}.}
    \label{fig:Fermi_PKS_2155-304}
\end{figure}

\begin{paracol}{2}
\switchcolumn

\vspace{-6pt}

\subsection{{Combined \textit{Fermi}-LAT and H.E.S.S. Observations of PKS~2155-304}}
Another study using the \textit{Fermi}-LAT data from the PKS 2155-304 was carried out by \linebreak \mbox{\citet{2018PhRvD..97f3009Z}}. The~ used data were taken from \textit{Fermi}-LAT observations in the energy range of 100~{MeV}--500~{GeV}. Photon--ALP oscillations in the inter-cluster magnetic field and the galactic magnetic field of the Milky Way are included. For~different sets of couplings in the range of $10^{-12}~\textrm{GeV}^{-1}\leq{g}_{a\gamma\gamma}\leq\times10^{-10}~\textrm{GeV}^{-1}$ and mass of ALPs $10^{-1}~\textrm{neV}\leq{m}_{a}\leq10^{2}~\textrm{neV}$ and 800 different realizations of the inter-cluster magnetic field, a~binned likelihood analysis similar to \citep{2016PhRvL.116p1101A} was performed. The~best fits with and without ALPs were compared to the observed spectrum, and the result is shown in Figure~\ref{fig:Fermi_PKS_2155-304}. A~joint likelihood was calculated; parameter space regions were excluded with $99.9\%$ confidence level and compared with the previous results from H.E.S.S.~\citep{2013PhRvD..88j2003A} and with the \textit{Fermi}-LAT observations of NGC~1275~\citep{2016PhRvL.116p1101A} in Figure~\ref{fig:Fermi_PKS_2155-304}.
\end{paracol}
\nointerlineskip

\begin{paracol}{2}
\switchcolumn

\vspace{-6pt}

\subsection{{\textit H.E.S.S. Study with Galactic Sources}}
More recently, H.E.S.S. data of galactic TeV $\gamma$-ray sources were used to search for  ALP oscillation effects \citep{2019JCAP...06..042L}. Ten sources, mainly supernova remnants and pulsar wind nebulae studied by H.E.S.S., were utilized. By~using sources in the galactic plane, one can probe the ALPs parameter space with higher ALP mass, $m_{a}>{10}^{-7}$~neV. This is due to the strength of the galactic magnetic field, an~important factor for the photon-ALP oscillation, as~seen in~Equation \eqref{eq:E_crit}. The~ALP model was obtained by multiplying a spectral fit without ALPs with the ${P}_{\gamma\gamma}$ for a certain parameter set ($m_{a},{g}_{a\gamma\gamma}$), and~including the instrument energy resolution. As~above, for~each set of parameters ($m_{a},{g}_{a\gamma\gamma}$) the ALP model was fitted to the observed spectrum, and  a~$\chi^{2}$ value was calculated and compared to the best fit over the whole parameter space. 
{The best parameters were deduced from the calculation of the $\chi^2$; however, as the photon--ALP conversion is degenerate in the coupling and magnetic field, and~that the induced irregularities are not linearly scaled with the ALPs parameters, a~threshold value for excluding the ALPs parameters was derived using the Monte Carlo simulations. In,~e.g.,~in~\citep{2019JCAP...06..042L} the threshold value is calculated from Monte Carlo simulations and compared to the difference in $\chi^{2}$ values for each set of ALPs parameters and the best fit over the whole parameter space. Since a scan of the whole parameter space is not feasible~\citep{2016PhRvL.116p1101A}, it is assumed that the overall shape probability distribution of the alternative hypothesis (with ALPs) can be approximated with the null distribution (no ALPs). It has been shown that such an approach yields conservative limits~\citep{2016PhRvL.116p1101A}.}
The results of~\citet{2019JCAP...06..042L} were consistent with other limits, but were uniquely sensitive towards the higher mass range. This showed that using galactic observations of TeV sources can improve and further constrain the high-mass part of the ALPs parameter~space. 

Other studies using \textit{Fermi}-LAT observations combined with IACTs results have been carried out, using the MAGIC~\citep{2018arXiv180504388M} and the H.E.S.S.~\citep{2020arXiv200207571G} data. In~\citep{2018arXiv180504388M}, both signatures induced by the photon--ALP oscillations and step-like flux suppression at the energies $E_{\gamma}>{E}_{crit}$ in the spectrum of NCG1275 were investigated. 
{
As can be seen, the~irregularity estimator in Equation~(\ref{eq:reduced_chi_square}) is the reduced-$\chi^{2}$. For~its general applicability in testing fits to the observed data, and~simplicity of calculation, the $\chi^{2}$ test has been used in several works~\citep{2018arXiv180504388M, 2019JCAP...06..042L}. For~each set of the considered ALPs and each magnetic field realization, and photon survival probability is calculated and multiplied by the best fit of the time-averaged spectrum, not including the ALPs effects. $\chi^{2}$ values for each of these fits are calculated. Testing of the ALPs hypothesis is performed using the $\Delta\chi^{2}$, defined as $\Delta\chi^{2}={\chi}^{2}_{wALP}-{\chi}^{2}_{w/oALP}$. Based on the distribution of these values for each set of ALPs parameters, the~exclusion region is evaluated under specific criteria and ALPs parameters are excluded.}
\citet{2018arXiv180504388M} considered 1000 different random realizations of the cluster magnetic field (modelled as in~\citep{2016PhRvL.116p1101A}) for a range of ALP parameters, coupling $10^{-14}~\textrm{GeV}^{-1}\leq{g}_{a\gamma\gamma}\leq10^{-9}~\textrm{GeV}^{-1}$ and mass of ALPs $10^{-2}~\textrm{neV}\leq{m}_{a}\leq10^{2}~\textrm{neV}$. By~combining observations of \textit{Fermi}--LAT, the MAGIC energy range available is extended, and  by observing both the patterns of the spectrum, a higher sensitivity to the photon-ALP coupling values is reached, dropping down to $g_{a\gamma\gamma}\sim10^{-12}~\textrm{GeV}^{-1}$.  
The result was the exclusion of the broader part of ALPs parameter space, compared to the previous analysis of the \textit{Fermi}-LAT data alone. The~excluded region also included the part of the ALPs parameter space which can be assigned to the possible ALP Dm. This showed the potential of combining data obtained by different instruments for the purpose of increasing part of the available ALPs parameter space and increasing the sensitivity.
Following previous interest in the effects ALPs oscillations could have on the BL~Lac spectra~\citep{2013A&A...554A..75S, 2013JCAP...11..023M}, recent works investigated the same using the simulations for the upcoming experiments and showed that BL~Lac could be used for future studies of the ALPs oscillations~\citep{2019MNRAS.487..123Gc}. 

\subsection{{Supernova Remnants}}
{Expanding their previous work,~\citep{2018PhRvD..97f3003X},~\citet{2019PhRvD.100l3004X} performed a search for spectral irregularities in three galactic supernova remnants, combining GeV data from Fermi-LAT and TeV data from IACTs (H.E.S.S., MAGIC and VERITAS). The~broadband spectra were fitted with models with and without photon--ALP conversion in the galactic magnetic field. The~ALP hypothesis was tested using the $\chi^2$ analysis and the combined limits were again shown to be inconsistent with limits already set by CAST. The~authors speculated that a possible reason for this result could be the uncertain connection between the \textit{Fermi}-LAT spectrum and the IACT observations, which are not easily calibrated in energy, and~also the systematic uncertainties of the instruments that were not taken into account~\citep{2019PhRvD.100l3004X}. This approach is likely to be revisited once CTA start taking data.}

\subsection{{Studies Obtained Comparing Data from Different Blazars}}
In~\citep{2020arXiv200207571G}, the \textit{Fermi}-LAT and H.E.S.S. data of two BL-Lacs are analyzed. Two different EBL models are also probed. {The ALP model included mixing of the inter-cluster magnetic field modeled as a Gaussian turbulent field with zero mean and variance~$\sigma_B$, as in~\citep{2014JCAP...09..003M}, and the Galactic magnetic field~\citep{2012ApJ...757...14J}. The~ALPs hypothesis was evaluated in a similar way, as in~\citep{2016PhRvL.116p1101A}, using a likelihood ratio test.} The results showed the improvement of the fit when ALP models are included and set constraints on the ALPs parameter space consistent with the previously obtained~ones. 

In~\citep{2020MNRAS.493.1553G} the highest energy spectra of AGN studied by \textit{Fermi}-LAT and IACTs are compared, showing that the inclusion  of proton--ALPs oscillation effects improves the agreement of the standard AGN model with the data.
Recently, the~analysis of~\mbox{\citet{2016PhRvL.116p1101A}} was revisited by~\citet{2020arXiv201012396C} using a different analysis method, calculating the irregularity of the spectrum of NGC~1275.
{Aiming to measure the irregularity of the spectrum, an~estimator needs to be chosen. Looking back to the article by H.E.S.S~\citep{2013PhRvD..88j2003A}, one could decide to use the estimator from Equation~(\ref{eq:estimator_hess}). A possible problem arises in the case of a large number of energy bins $(\sim$100) (as in~\citep{2020arXiv201012396C}), since the ALPs signatures might become wider than the bin size, making this kind of estimator insensitive to such alternations. Using the energy windows, instead of the spectral points triplets, and~following the assumption of a power-law model in those energy windows, \citet{2020arXiv201012396C} proposed an alternative version of the estimator,
\begin{equation}
    {\mathcal{I}}_{alt}=\sum_{i}\sum_{j}\frac{{\big({\phi}_{i,j}^{pl}-{\phi}_{i,j}\big)}^{2}}{\sigma_{i,j}^{2}}.
    \label{eq:alternated_estimator}
\end{equation}
where $i$ and $j$ represent the energy window and bin, respectively, while ${\phi}^{pl}$ is the flux assumed by the power-law spectral fit in each energy window, and~$\phi$ and $\sigma$ are the measured values of the flux and uncertainty, respectively~\citep{2020arXiv201012396C}. Each of the simulated ALPs models were fitted assuming a baseline log parabola. From~the assumption that the observed irregularity can be explained by the photon conversion connected to a given set of ALPs parameters, exclusion limits were set.} 
This study included mixing in the intracluster magnetic field and in the galactic magnetic field of the Milky Way. Excluded couplings are ${g}_{a\gamma\gamma}>3\times{10}^{-12}~\textrm{GeV}^{-1}$ for masses of the ALPs ${m}_{a}\sim1$~neV at a $95~\%$ confidence level. The~results of this search show the possibility of further improving the constraints by combining NGC~1275 observations with observations of another source PKS~2155-304~\citep{2020arXiv201012396C}. 


\subsection{{ALP-Photon Back Conversion in the Galactic Magnetic Field }}
As investigated by~\citet{2021arXiv210110270L}, new
observations of VHE $\gamma$-ray sources could lead to the detection of the flux enhancement due to the ALP--photon back-conversion in the Galactic magnetic field. This enhancement is expected at energies $E_{crit}\sim100$~TeV~\citep{2021arXiv210110270L} and could be detected by the Large High-Altitude Air Shower Observatory (LHAASO)~\citep{2019arXiv190502773B}, CTA, and~by the planned Southern Wide-field Gamma-ray Observatory (SWGO)~\cite{https://doi.org/10.1002/asna.202113946}. \mbox{\citet{2021arXiv210110270L}} analyzed HE and VHE $\gamma$-ray data from three promising AGNs and the spectra were extrapolated to the energies $E\sim100$~TeV. Further on, the~assumed intrinsic spectra were folded with the $P_{\gamma\gamma}$, assuming the photon--ALP conversions in the source magnetic field and the back-conversion in the magnetic field of the Milky Way. These spectra were compared to the ones obtained only by including the EBL and CMB attenuation. The~results showed that, in the respective energy range (above $E\sim100$~TeV), predicted flux enhancement is above one order of magnitude and higher than the sensitivity of the instrument, which will allow for the constraints to be set on the ALPs parameter space~\citep{2021arXiv210110270L}. It is also emphasized that, to set stringent constraints, a~better estimation of the intrinsic spectra, magnetic fields and EBL attenuation needs to be established, all of which are expected with the upcoming experiments in HE and VHE $\gamma$-ray~astronomy. 

The next generation of IACTs, CTA, is expected to lower the uncertainties in the spectra of astrophysical sources such as active galaxies and BL Lacs and increase the sensitivity to photon--ALP oscillations~\citep{Abdalla_2021}. In~that way, they may surpass the current constraints by broadening the part of the ALPs parameter space available for $\gamma$-ray studies, and excluding it even further.
A comparable performance is expected from future laboratory experiments: the~axion helioscope IAXO~\citep{2015PhPro..61..193V} and Any Light Particle Search II (ALPSII)~\citep{2016arXiv161105863S}.

\section{Outlook: The Cherenkov Telescope~Array}
\label{outlook}
In the previous section, we discussed the current constraints set by IACTs on ALPs. As~noted, to date, several studies using H.E.S.S. and MAGIC data have been performed, but~there is still room for improvement with the new upcoming generation of experiments. In~particular, CTA is expected to probe the energies up to $E_{\gamma}\approx300$~TeV, which directly improves the possibility of studying ALPs manifestations. \citet{Abdalla_2021} created simulations of the observation of the radio galaxy NGC~1275. The~magnetic field of the Perseus cluster is modeled following \citet{2012ApJ...757...14J}, with morphology modeled as a random field with Gaussian turbulence. The~conservative value of the central magnetic-field strength was set to 10 $\upmu$G, along with the other parameters listed in~\citep{Abdalla_2021}. Using three different sets of ALPs parameters with 100 different magnetic field realizations, $P_{\gamma\gamma}$ was calculated using the GAMMAALPs code\footnote{\url{https://github.com/me- manu/gammaALPs}  (accessed on June, 3rd 2020).}. GAMMAALPs solves the equations of motion for the photon--ALP system using the transfer matrix method. Considering other effects that could impact the photon flux, GAMMAALPs includes the EBL absorption, dissipation in QED, and~CMB effects. Observations in both the quiescent and the flaring state are included in a $\approx$300~h exposure. The authors included the systematic uncertainties of the instrument, and  fits are performed both with and without  ALPs effects.  As the energy binning has a great importance for observing wiggles in the spectrum, three different sets of parameters are used, and fits for each of them are performed  by maximizing the likelihood and summing over $40$ energy bins. For~each set, $100$ different magnetic field realizations are computed and likelihood values corresponding to quantile $Q=0.95$ of the distributions are chosen. To~obtain the confidence intervals of $95\%$ and $99\%$, Monte Carlo simulations are~used. 

The results showed that, in contrast to the quiescent state, the~flaring state of the source provides a stronger exclusion of the ALPs parameter space, reaching a level 
where ALPs could constitute the entirety of DM. A~probable reason for this is a strong background cut on the quiescent data, which causes the exclusion of the low energy bins from the analysis. On~the other side, flaring state observations extend to lower energies. As~concluded by the authors, this shows the great importance of observing the high activity states of this and other sources that have yet to be studied. Changes in the magnetic field parameters are also tested and the projected exclusion parameters are presented in Figure~\ref{fig:projected_exclusions_CTA}.

%
It is important to note  that the constraints on the ALPs parameter space are sensitive to changes in the assumed parameters' values in the model of the magnetic field of the Perseus galaxy cluster. Moreover, it is found that finer energy binning gives stronger constraints, as the analysis becomes more sensitive to small and fast oscillations in the spectrum, caused by the photon--ALP oscillations. 
The projected limits obtained in~\citep{Abdalla_2021} can be seen in Figure~\ref{fig:projected_exclusions_CTA}. Compared to future laboratory experiments (e.g. ALPSII \citep{2016arXiv161105863S}), CTA exclusions of the ALPs parameter space will be dominated by the systematic uncertainties of the model \citep{Abdalla_2021}; CTA is expected to have a similar sensitivity to the planned IAXO~\citep{2015PhPro..61..193V} and ALPS II experiments~\citep{2016arXiv161105863S}. 
\begin{figure}[H]
    \includegraphics[width=0.8\linewidth]{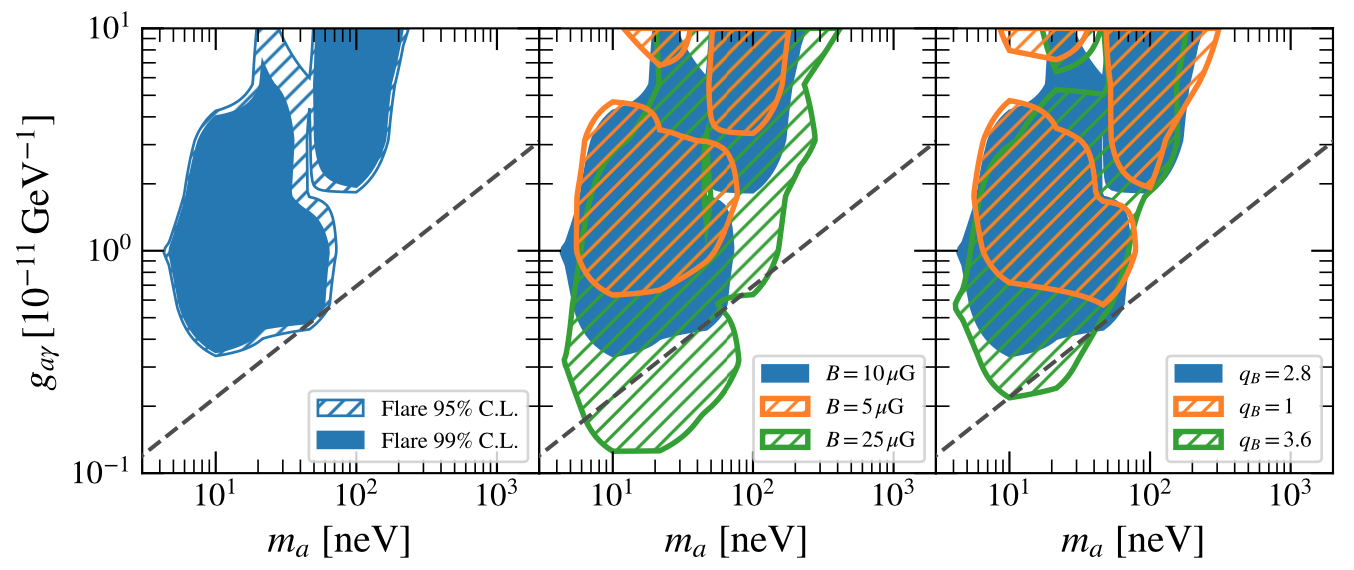}
    \includegraphics[width=0.6\linewidth]{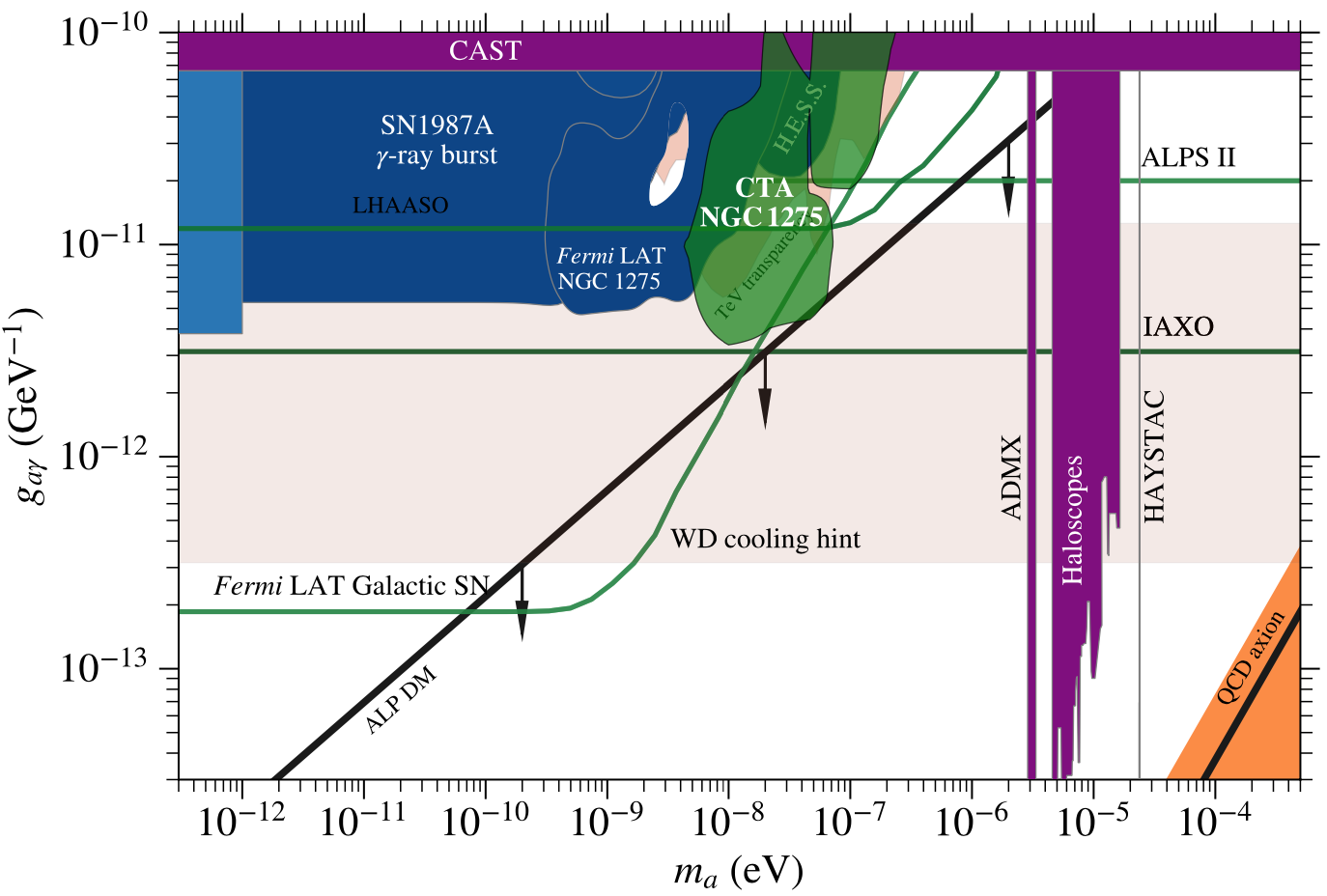}
    \caption{(\textbf{top}) Projected CTA exclusions on the ALPs parameter space for different assumptions on the intracluster magnetic field parameters. Reprinted from~\citet{Abdalla_2021}  (\textbf{bottom}) Projected limits from the CTA simulations, compared to constraints on the ALPs parameter space with \textit{Fermi} LAT and H.E.S.S. Reprinted from~\citet{Abdalla_2021}.}
    \label{fig:projected_exclusions_CTA}
\end{figure}

\section{Summary and~Conclusions}
ALPs are one of the most promising candidates to solvethe strong $CP$ problem, and~also a viable solution to the long-lived mystery of astrophysics, the~existence of  DM. 
In this review, searches for ALPs are presented, focusing on VHE $\gamma$-ray astronomy and IACTs. Current constraints set with IACTs~\citep{2013PhRvD..88j2003A} are still viable, and~complementing constraints are set by other experiments and instruments, such as axion helioscopes,~see, e.g., \citep{2010PhLB..689..149E, PhysRevD.92.092002, 2016PhRvL.116p1101A, 2017NatPh..13..584A, 2018PhRvL.120o1301D}. Even though a great number of searches have been performed, ALPs parameter space still leaves room for future developments. Probably the most interesting part of yet-unexplored ALPs parameter space is accounting for ALPs  which are able to explain and constitute most or all of the DM in the universe. This region is anticipated in the future $\gamma$-ray experiments, such as CTA~\citep{Abdalla_2021} and SWGO~\cite{https://doi.org/10.1002/asna.202113946}, or~in LHAASO~\citep{2021arXiv210110270L}, that will be able to explore higher energies of up to about $100$ TeV and exclude masses of ALPs of ${m}_{a}\sim\,200$~neV~\citep{Abdalla_2021}. With~these and other upcoming laboratory axion experiments, constraints on the ALPs parameter space, or even a possible detection of the ALPs, are increasingly~anticipated. 




\vspace{6pt} 



\authorcontributions{Conceptualization, I.B, M.D; methodology, I.B.; validation, A.D.A; writing---original draft preparation, I.B, M.D. and M.M.; writing---review and editing, A.D.A, M.M.; supervision, M.D.; All authors have read and agreed to the published version of the manuscript.}

\funding{This research received funding from the Italian Ministry of Education, University and Research (MIUR) through the ``Dipartimenti di eccellenza” project Science of the Universe, the University of Padova for the XXXVIII PhD call cycles. M.M. acknowledges the Croatian Science Foundation (HrZZ) Project IP-2016-06-9782 and the University of Rijeka Project 13.12.1.3.02.}


\conflictsofinterest{The authors declare no conflict of interest. The~funders had no role in the design of the study; in the collection, analyses, or~interpretation of data; in the writing of the manuscript, or~in the decision to publish the~results.} 

 
\clearpage
\abbreviations{Abbreviations}{The following abbreviations are used in this manuscript:\\

\noindent 
\begin{tabular}{@{}ll}
ADMX & Axion Dark Matter eXperiment \\
AGN & Active Galactic Nucleus \\
ALPs & Axion-like particles \\
ALPS & Any Light Particle Search \\
ATNF & The Australia Telescope National Facility \\
CAST & The CERN Axion Solar Telescope \\
CERN & European Council for Nuclear Research \\
CDM & Cold Dark Matter \\
CMB & Cosmic Microwave Background \\
CP & Charge-Parity \\
DAMA & Dark Matter experiment \\
DARMA & De Angelis, Roncadelli and Mansutti \\
DFSZ & Dine--Fischler--Srednicki--Zhitnitsky \\
DM & Dark Matter \\
EBL & Extragalactic Background Light \\
FSRQ & Flat Spectrum Radio Quasar \\
GMF & Galactic Magnetic Field \\
HE & high-energy (E > 100~MeV) \\
H.E.S.S. & The High-Energy Stereoscopic System \\
IACT & Imaging Atmospheric Cherenkov Telescope \\
IAXO & The International Axion Observatory \\
ICMF & Intracluster Magnetic Field \\
IGMF & Intergalactic Magnetic Field \\
KSVZ & Kim--Shifman--Vainshtein--Zakharov \\
LHAASO & The Large High-Altitude Air Shower Observatory \\
LHC & Large Hadron Collider \\
LIV & Lorentz Invariance Violation \\
MAGIC & Major Atmospheric Gamma-ray Cherenkov \\
OSQAR & The Optical Search for QED Vacuum Bifringence \\
PQ & Peccei-Quinn \\
PVLAS & The Polarization of the Vacuum with Laser \\
QCD & Quantum Chromo-Dynamics \\
SM & Standard Model \\
SWGO & The Southern Wide-field Gamma-ray Observatory \\
VERITAS & Very Energetic Radiation Imaging Telescope Array System \\
VHE & very-high-energy (E > 100~GeV) \\
WISPs & Weakly Interacting Slim Particles \\

\end{tabular}}

\end{paracol}
\reftitle{References}

\end{document}